\documentclass[fleqn,usenatbib]{mnras}

\usepackage[T1]{fontenc}

\usepackage{graphicx}	
\usepackage{amsmath}	
\usepackage{amssymb}	
\usepackage{hyperref}

\title[VLA Survey of M\,33]{Cloud-scale Radio Surveys of Star Formation and Feedback in Triangulum Galaxy M\,33: VLA Observations}

\author[F.~S. Tabatabaei et al.]{
F.~S. Tabatabaei,$^{1,2,3,4}$\thanks{E-mail: ftaba@iac.es }
W. Cotton$^{5}$,
E. Schinnerer$^{3}$,
R. Beck$^{4}$,
A. Brunthaler$^{4}$,
\newauthor
K.~M. Menten$^{4}$,
J. Braine$^{6}$,
E. Corbelli$^{7}$,
C. Kramer$^{8}$,
J. E. Beckman$^{2,9}$,
\newauthor
J. H. Knapen$^{2,9}$,
R. Paladino$^{10}$,
E. Koch$^{11}$,
and A. Camps Fari\~{n}a$^{2,9,12}$
\\
$^{1}$School of Astronomy, Institute for Research in Fundamental Sciences (IPM), PO Box 19395-5531, Tehran, Iran\\
$^{2}$Instituto de Astrof\'{i}sica de Canarias, V\'{i}a L\'{a}ctea S/N, 38205 La Laguna, Spain\\
$^{3}$Max-Planck-Institut f\"ur Astronomie, K\"onigstuhl 17, 69117, Heidelberg, Germany\\
$^{4}$Max-Planck Institut f\"ur Radioastronomie, Auf dem H\"ugel 69, 53121 Bonn, Germany\\
$^{5}$National Radio Astronomy Observatory (NRAO), 520 Edgemont Road, Charlottesville, VA 22903, USA\\
$^{6}$Laboratoire d'Astrophysique de Bordeaux, Univ. Bordeaux, CNRS, B18N, All\'{e}e Geoffroy Saint-Hilaire, 33615 Pessac, France\\
$^{7}$INAF-Osservatorio Astrofisico di Arcetri, Largo E. Fermi, 5, 50125 Firenze, Italy\\
$^{8}$Institut   de   Radioastronomie   Millim\'{e}trique   (IRAM),   300   Rue de   la   Piscine,   38406   Saint   Martin   d'H\`{e}res, France\\
$^{9}$Departamento de Astrof\'{i}sica, Universidad de La Laguna, 38206 La Laguna, Spain\\
$^{10}$INAF - Istituto di Radioastronomia, via P. Gobetti 101, 40129 Bologna, Italy\\
$^{11}$Center for Astrophysics, Harvard \& Smithsonian, 60 Garden St., Cambridge, MA 02138, USA\\
$^{12}$Departamento de la Tierra y Astrof\'{i}sica, Facultad de CC F\'{i}sicas, Universidad Complutense de Madrid, 28040 Madrid, Spain
}

\date{Accepted XXX. Received YYY; in original form ZZZ}

\pubyear{2019}

\begin{document}
\label{firstpage}
\pagerange{\pageref{firstpage}--\pageref{lastpage}}
\maketitle

\begin{abstract}
Studying the interplay between massive star formation and the interstellar medium (ISM) is paramount to understand the evolution of galaxies. Radio continuum (RC) emission serves as an extinction-free tracer of both massive star formation and the energetic components of the interstellar medium. We present a multi-band radio continuum survey of the local group galaxy M~33 down to $\simeq$30\,pc linear resolution observed with the Karl G. Jansky Very Large Array (VLA).  We calibrate the star-formation rate surface density and investigate the impact of diffuse emission on this calibration using a structural decomposition. Separating the thermal and nonthermal emission components, the correlation between different phases of the interstellar medium and the impact of massive star formation are also investigated. Radio sources {with sizes~$\lesssim$~200~pc} constitute about 36\% (46\%) of the total RC emission at 1.5~GHz (6.3~GHz) in the inner 18\arcmin $\times$18\arcmin~(or 4~kpc~$\times$~4~kpc) disk of M~33.  The nonthermal spectral index becomes flatter with increasing star-formation rate surface density, indicating the escape of cosmic ray electrons {from their birth places}. The magnetic field strength also increases with star-formation rate following a bi-modal relation, indicating that the small-scale turbulent dynamo acts more efficiently at higher luminosities and star-formation rates. Although the correlations are tighter in star-forming regions, the nonthermal emission is correlated also with the more quiescent molecular gas in the ISM. An almost linear molecular star-formation law exists in M\,33 when excluding diffuse structures. Massive star formation amplifies the magnetic field and increases the number of high-energy cosmic ray electrons, which can help the onset of winds and outflows.

\end{abstract}




\section{Introduction}
The RC emission from galaxies is mainly linked to their massive star formation and nuclear activity  \citep[e.g.,][]{Condon_92}. Massive stars produce RC emission while in their early phases of evolution (young stellar objects such as protostars with radio jets/outflows and young O and early B-type stars creating HII regions) and in their late phases of evolution (supernova explosions, SNe, and supernova remnants, SNRs). Hence, being dust-unbiased, RC surveys are ideal tools to trace different evolutionary phases of massive stars. Apart from tracing these stars, the way they interact with the interstellar medium (ISM) and the role of the ISM in the star formation process
are key questions that address the evolution of galaxies \citep[][]{Taba_18}. Resolved and sensitive RC observations of nearby galaxies exhibit extended structures besides bright star-forming regions, indicative of the diffuse ionized, magnetized, and relativistic ISM \citep[e.g.,][]{Beck_15, Krause_18,For_18}. These RC maps are ideal to address the effect of massive star formation on the ISM and vice versa.

Careful separation of the thermal and nonthermal RC components in galaxies shows that the  cosmic ray electrons (CRes), which have a relatively flat power-law spectral index, are more energetic in complexes of star-forming regions, where the magnetic field is also stronger \citep{Tabatabaei_3_07,Taba_13,Hassani}. Hence, massive star formation can insert strong nonthermal pressure into the ISM through injection of CRes and the amplification of  magnetic fields, besides their thermal feedback. This can cause cosmic ray-driven winds and outflows in galaxies 
\citep[][]{Taba_17}. Understanding the formation of the next generation of stars in such a magnetized and turbulent ISM is a pressing question. Sensitive RC observations on cloud scales in nearby galaxies are vital to study the formation and regulation of stars in the presence of magnetic fields and cosmic rays.

\begin{table}
\begin{center}
\caption{Positional data on M\,33.}
\begin{tabular}{ l l }
\hline
\hline
Nucleus position\,(J2000)$^{1}$    & RA\,=\,$1^{h}33^{m}51.0^{s}$      \\
    &  DEC\,=\,$30^{\circ}39\arcmin37.0\arcsec$\\
Position angle of major axis$^{2}$   &23$^{\circ}$ \\
Inclination$^{3}$    & 56$^{\circ}$ \\
Distance$^{4}$\,(1$\arcsec$=\,3.5\,pc)   & 730\,kpc\\
\hline
\noalign {\medskip}
\multicolumn{2}{l}{$^{1}$ \cite{devaucouleurs_81}}\\
\multicolumn{2}{l}{$^{2}$ \cite{Deul}}\\
\multicolumn{2}{l}{$^{3}$ \cite{Regan_etal_94}}\\
\multicolumn{2}{l}{$^{4}$ \cite{Brunthaler}}\\
\end{tabular}
\label{tab:m33}
\end{center}
\end{table}
In normal star-forming galaxies, a linear correlation is expected between the star-formation rate (SFR) and the  RC  emission  if the  production rate of CRes, assumed to be proportional to the SNe  rate, is proportional  to the SFR and if CRes lose their energy before escaping from a galaxy \citep{Voelk, Condon_92}. A detailed study of the mid radio wavelength spectral energy distribution (1-10\,GHz SED) showed that while the correlation is linear with the thermal component, it  deviates from linearity for the nonthermal component \citep[galaxies with higher SFR are brighter in nonthermal RC,][]{Taba_17}. A similar deviation from linearity had also been reported {in integrated studies \citep[e.g.,][]{Price,Niklas_977,Gurkan,smith21}}. This can be expected as the synchrotron emission depends not only on the CRe number but also the magnetic field strength. Galaxies with higher SFR also have stronger magnetic fields \citep[e.g.,][]{chyzy_11,Heesen_14,Taba_17}, likely due to feedback from star formation \citep{Schleicher}. To understand the origin of this non-linearity, it is hence vital to resolve details of the connection of SFR with both the CRes (number and energy) and the magnetic field (strength and structure).

Studies of the RC correlation with the SFR surface density ($\Sigma_{\rm SFR}$) or the resolved RC--IR correlation show that  different relations hold depending on galactic region, e.g. star-forming vs. non-star-forming \citep{Taba_13} or spiral arms vs. inter-arm regions \citep{Dumas}, and also on spatial scales \citep[e.g.,][]{Taba_13_b}. Taking all structures together, \cite{Heesen_14} found a sub-linear RC vs. $\Sigma_{\rm SFR}$ relation (or super-linear $\Sigma_{\rm SFR}$ vs. RC) in a number of galaxies studied in the SIRTF Nearby Galaxies Survey \citep[SINGS,][]{Kennicutt_03}, concluding that using the RC emission alone is not sufficient to measure $\Sigma_{\rm SFR}$ locally. However, averaging over different galactic structures can also impose a bias on the $\Sigma_{\rm SFR}$ measurements due to the inclusion of emission unrelated to the current star-formation activity. Decomposition of emission emerging from different galactic components such as star-forming regions, extended structures, and diffuse disk(+halo), i.e., structural decomposition, can hence be important when measuring the SFR using resolved RC maps.

Due to its proximity \citep[$d=$730 kpc, 1$\arcsec \simeq$ 4\,pc, ][]{Freedman_etal_91} and viewing angle \citep[$i=56^{\circ}$,][]{Regan_etal_94}, the Triangulum Galaxy M\,33 (NGC\,598) provides a unique laboratory to resolve the ISM structures and dominant processes in both star-forming and non-star-forming regions. Radio surveys of M\,33 have mainly targeted large-scale ISM structures \citep[$\gtrsim 1$\,kpc,][]{Israel_74,vonkap,Beck_79,Buczilowski_87} or bright sources such as HII regions and SNRs  \citep{viallefond_et_al_98,Duric,White}. In the latter surveys, the ISM adjacent to bright radio sources is usually neglected due to either insufficient sensitivity or the missing short spacing problem in radio interferometry. Recovering the short-spacing emission using an image reconstruction technique allowed \cite{Tabatabaei_1_07,Tabatabaei_2_07,Taba_13_b} to study M\,33's ISM structures down to 200-pc scales (50\arcsec). Here, we report a new wide-band, full polarization RC survey of M\,33 with the {\it Karl G. Jansky Very Large Array} (VLA) at C (6.3\,GHz central frequency) and L (1.5\,GHz central frequency) bands. These observations allow us to study the ISM down to 30-pc scales addressing not only star-forming regions but also the more quiescent, non-star-forming regions which are needed to investigate the regulation of star formation in this galaxy.

Taking advantage of the VLA's wide-bandwidth WIDAR correlator, we present resolved RC maps of M\,33 at significantly higher sensitivity than before \citep[][]{Tabatabaei_1_07,Tabatabaei_2_07}. The VLA observations were designed to achieve a resolution similar to those of the IRAM 30m CO(2-1) \citep{Gratier,Druard} and the Herschel PACS continuum observations \citep{Kramer10,Boquien_15} to perform a consistent comparison of the various ISM phases and components.

Thanks to these high-resolution and -sensitivity data, this paper presents a study of 1) the effect of diffuse galactic components calibrating the $\Sigma_{\rm SFR}$, 2) the correlation between RC-emitting components and neutral gas phases of the ISM and 3) the impact of massive star formation on relativistic and magnetized ISM.

The paper is organized as follows: After describing the observations and data in Sect.~2 {\footnote{The polarization study will be presented elsewhere.}} and the thermal/nonthermal separation method (Sect.~3), we present the resulting  observations, as well as the thermal and nonthermal maps, the spectral index, and equipartition magnetic field  in Sect.~4. Then we discuss the SFR--RC calibrations,  the RC--neutral gas correlations, and investigate the impact of star formation on CRes and magnetic fields (Sect.~5). Our findings are summarized in Sect.~6.

\section{Data}

\subsection{VLA Observations and Data Reduction}

We performed interferometer observations with the Karl G. Jansky Very Large Array (VLA\footnote{The VLA is a facility of the National Radio Astronomy Observatory. The NRAO is operated by Associated Universities, Inc., under contract with the National Science Foundation.}).
The VLA observations were taken using the wideband, spectral line correlator WIDAR in full polarization under proposal number 11B-145 (PI: F. Tabatabaei). The observations were carried out in D-configuration in December 2011 and C- configuration in April 2012 at C (5.5-7.5\,GHz) and L (1-2\,GHz) bands. The corresponding scheduling blocks (SBs) and dates of observations used in this study are listed in Table~\ref{tab:date}.
\begin{table}
\begin{center}
\caption{VLA observations, scheduling blocks (SBs), dates, and total observation times }
\begin{tabular}{llll}
\hline
Band & SB ID &  Observation Date & Total time (hr) \\ \hline \hline
C& 5145344 & 5 December 2011 & 5\\
C& 5765541 & 6 December 2011 & 5\\
C& 5758725 & 12,13 December 2011 & 5\\
C& 5760250 & 13,14 December 2011 & 5\\
C& 5767346 & 17,18,21 December 2011 & 10\\
L& 9644094 & 21 April 2012 & 5.5 \\
\hline
\end{tabular}
\label{tab:date}
\end{center}
\end{table}

At C band, two continuous base-bands, each with 8 sub-bands of 128\,MHz width, were tuned to cover 5--5.9\,GHz and
6.7--7.6\,GHz. These observations image the central $18'\times18'$ area including the central extended region and the main spiral arms.
As the primary beam FWHM of $6'$ (at 7.5\,GHz) is much smaller than this area, we performed mosaicing.
Following \cite{Condon_98}, a distance between pointings of $\theta_{\rm pb}/ \sqrt{2} \simeq 4'$ was needed to cover this area to obtain a composite image with uniform   sensitivity leading to a mosaic of 25 pointings (5 pointings along each RA and DEC).

The L-band (1--2\,GHz) observations were carried out in C configuration. We used two base-bands of 512\,MHz at their default frequencies of 1.25\,GHz and 1.75\,GHz, each with 8 sub-bands of 64\,MHz.
A mosaic of 9 pointings with a pointing separation of $20'$ covered a $50'\times 50'$ area.

The sources 3C\,48, 3C\,138 were observed for calibrating the flux density and polarization angle.

\begin{table}
\begin{center}
\caption{Central positions of the pointings observed with VLA at C (5--7.6\,GHz) and L (1--2\,GHz) bands.}
\begin{tabular}{ l l l l}
\hline

 Band & Pointing &   R.A.\, &   DEC.\, \\
       & $\#$   &(J2000) &(J2000) \\
       &   &   ($^h$ $^m$ $^s$) & ($^{\circ}$ $\arcmin$ $\arcsec$)\\
 \hline
\hline
C&1  & 01 34 40.00 & +30 48  5.80  \\
C&2  & 01 34 20.27 & +30 48  5.80 \\
C&3  & 01 34  0.54 & +30 48  5.80  \\
C&4  & 01 33 40.82 & +30 48  5.80 \\
C&5  & 01 33 21.09 & +30 48  5.80 \\
C&6  & 01 34 40.00 & +30 43 51.40 \\
C&7  & 01 34 20.27 & +30 43 51.40 \\
C&8  & 01 34  0.54 & +30 43 51.40 \\
C&9  & 01 33 40.82 & +30 43 51.40 \\
C&10 & 01 33 21.09 & +30 43 51.40 \\
C&11 & 01 34 35.00 & +30 39 37.00 \\
C&12 & 01 34 15.29 & +30 39 37.00 \\
C&13 & 01 33 55.57 & +30 39 37.00  \\
C&14 & 01 33 35.86 & +30 39 37.00  \\
C&15 & 01 33 16.15 & +30 39 37.00  \\
C&16 & 01 34 34.00 & +30 35 22.60 \\
C&17 & 01 34 14.30 & +30 35 22.60 \\
C&18 & 01 33 54.60 & +30 35 22.60  \\
C&19 & 01 33 34.90 & +30 35 22.60  \\
C&20 & 01 33 15.20 & +30 35 22.60 \\
C&21 & 01 34 30.00 & +30 31  8.20 \\
C&22 & 01 34 10.31 & +30 31  8.20 \\
C&23 & 01 33 50.63 & +30 31  8.20 \\
C&24 & 01 33 30.95 & +30 31  8.20 \\
C&25 & 01 33 11.26 & +30 31  8.20 \\
\hline
L&1  & 01 35 11.00 & +30 19  37.00  \\
L&2  & 01 33 51.00 & +30 19  37.00 \\
L&3  & 01 32 31.00 & +30 19  37.00 \\
L&4  & 01 32 31.00 & +30 39  37.00 \\
L&5  & 01 33 51.00 & +30 39  37.00\\
L&6  & 01 35 11.00 & +30 39  37.00\\
L&7  & 01 35 11.00 & +30 59  37.00\\
L&8  & 01 33 51.00 & +30 59  37.00 \\
L&9  & 01 32 31.00 & +30 59  37.00\\
\hline
\hline
\end{tabular}
\label{tab:pos}
\end{center}
\end{table}

The data were processed in the Obit package \citep{Cotton}.  Calibration used the standard VLA calibration pipeline procedure.  Initial flagging used a comparison with running
medians in time and frequency to detect outliers. The switched power
system was used to determine short-term variations in the receiver
gain.  Then group delay, bandpass and amplitude and phase calibration
were determined using 3C48. Outliers among the calibration solutions
were used to flag the data.  After further flagging in frequency using
a running median, the calibration steps were repeated.  Polarization
calibration used 3C\,48 and 3C\,138 with the polarization model of \cite{Perley_13}.
The nonlinear fitting did not fix any of the instrumental polarization parameters.

Imaging of each pointing in Stokes I, Q and U used the wideband
Obit imager MFImage \citep{Cotton_18}.  Variations in
sky brightness and antenna gain with frequency were accommodated
using 5\% fractional bandwidth sub-bands which are imaged
independently but deconvolved jointly.
A multiresolution CLEAN (roughly 9\arcsec, 31\arcsec, and 57\arcsec)
was used to recover a wide range of size scales.
Then those pointings with a peak over 3 mJy/beam were also phase self-calibrated.  The calibrator was imaged in the same way as the targets which resulted in some imaging artifacts being incorporated into the self-calibration model (and preserved). Briggs Robust image weighting
resulted in the beams of all pointings being about 9\arcsec.2 $\times$ 8\arcsec.8 in C
band and 12\arcsec.7 $\times$ 12\arcsec.5 in L band. Each pointing image was then convolved to a circular beam before combination to get consistent resolution (9\arcsec.35 and 15\arcsec~in C- and L-band, respectively).

   A mosaic of the field for each image plane and for each frequency
sub-band and Stokes parameter was derived by the summation over
overlapping pointing images:
$$ M(x,y)\ =\ {{\sum_{i=1}^{n}A_{i}(x,y)\ I_{i}(x,y)}\over{\sum_{i=1}^{n}A^2_{i}(x,y)}},$$
where $A_{i}(x,y)$ is the antenna gain of pointing $i$ in
direction $(x,y)$, and $I_{i}(x,y)$ is the pointing $i$ pixel value
interpolated to direction $(x,y)$ and $M$ is the mosaiced image.

The rms noise at each spectral window is given in Table~\ref{tab:rms}.
At C-band, combining the spectral window images led to a final rms noise of 6\,$\mu$Jy per 9.35\arcsec~beam width at the central frequency of 6.3\,GHz that is close to the thermal noise. Broadband radio frequency interfence (RFI) affected the L-band observations particularly at the sub-bands 1, 2, and 10 leading to relatively higher noise level (sub-band 10 was totally flagged). The rms noise of the combined L-band spectral windows is about 37\,$\mu$Jy per 15\arcsec~beam width at 1.5\,GHz.

\begin{table}
\begin{center}
\caption{Rms noise in the sub-band images at 9.35\arcsec (C-band) and 15\arcsec (L-band) resolutions.}
\begin{tabular}{ l l l l l}
\hline
       & C-Band & \,\,\,\, \,\,\,\,\,\,\,\,  & \,\,\,\, \,\,\,\,\,\,\,\,L-Band& \\
\hline
Sub-band   & Central &   rms  &\,\,\,\, \,\,\,\,\,\,\,\,    Central & rms  \\
 \,\,\,\,\,\, $\#$    &  frequency   & noise & \,\,\,\, \,\, frequency   & noise \\
    & (MHz)& ($\mu$Jy/  &\,\,\,\, \,\,\,\,\,\,\,\,      (MHz)& ($\mu$Jy/\\
      & & beam) &\,\,\,\, \,\,\,\,\,\,\,\,      & beam)\\
\hline
\hline
1& 5050.4 & 35 &\,\,\,\, \,\,\,\,\,\,\,\,  1024.5 &360 \\
2& 5178.4 & 25 &\,\,\,\, \,\,\,\,\,\,\,\, 1088.5  &350\\
3& 5306.4 & 25 &\,\,\,\, \,\,\,\,\,\,\,\,  1152.5&200\\
4& 5434.4 & 25 &\,\,\,\, \,\,\,\,\,\,\,\, 1216.5 & 216\\
5& 5562.4 & 24 &\,\,\,\, \,\,\,\,\,\,\,\, 1280.5 &122\\
6& 5690.4 & 25 &\,\,\,\, \,\,\,\,\,\,\,\,  1344.5& 88\\
7& 5818.4 & 26 &\,\,\,\, \,\,\,\,\,\,\,\,  1408.5 &90\\
8& 5946.4 & 26 &\,\,\,\, \,\,\,\,\,\,\,\, 1472.5& 102\\
9& 6693.4 & 27 &\,\,\,\, \,\,\,\,\,\,\,\, 1524.5& 151\\
10& 6821.4 & 22 &\,\,\,\, \,\,\,\,\,\,\,\, 1588.5 & ...\\
11& 6949.4 & 24 &\,\,\,\, \,\,\,\,\,\,\,\,  1652.5& 107 \\
12& 7077.4 & 23 &\,\,\,\, \,\,\,\,\,\,\,\,  1716.5& 102\\
13& 7205.4 & 23 &\,\,\,\, \,\,\,\,\,\,\,\, 1780.5& 110\\
14& 7333.4 & 26  &\,\,\,\, \,\,\,\,\,\,\,\, 1844.5 & 112\\
15& 7461.4 & 25  &\,\,\,\, \,\,\,\,\,\,\,\, 1908.5 &120\\
16& 7589.4 & 27  &\,\,\,\, \,\,\,\,\,\,\,\, 1972.5 & 150\\
1-16 & 6319.9 & 6 &\,\,\,\, \,\,\,\,\,\,\,\,  1498.5 & 37\\
\end{tabular}
\label{tab:rms}
\end{center}
\end{table}

\subsection{Other Data}
To increase the image fidelity at 1.5\,GHz, we also used the VLA L-band data taken in the D-configuration and presented by \cite{Tabatabaei_2_07}. The 100-m Effelsberg maps at 6\,cm \citep{Tabatabaei_2_07} and 20\,cm \citep{Fletcher} were used to correct for the missing short spacing of the VLA data.

H$\alpha$ line emission was mapped with the 0.6 meter Burrell--Schmidt telescope at the Kitt Peak National Observatory by \cite{Hoopes_et_al_97H}, covering a $68\arcmin$\,$\times$\,$68\arcmin$ field of view. We corrected the data for the neighbouring NII line emission using the NII/H$\alpha$--M$_{\rm B}$ relation given by \cite{kennicutt_08}. We also used H$\alpha$ velocity dispersion data of bright HII regions taken with the  scanning Fabry-Perot interferometer on the 1.6m telescope at the Observatoire de Mont M\'{e}gantic (OMM, Quebec) in September 2012. Observing details can be found in \cite{kam}. The reduction and calibration  of the data cube was performed using the FPReduc package \citep{Diagle}. This pipeline provides tools to calibrate automatically Fabry-Perot data into velocity units, detects, and subtracts the continuum from the emission lne, and masks out the areas of the galaxy which do not have line emission. 

M\,33 was observed with the MIPS instrument \citep{Rieke} onboard the Spitzer Space Telescope at 24$\mu$m. The basic data reduction and mosaicing were performed with the  MIPS  instrument  Data  Analysis  Tool  version  2.90 \citep{Gordon_05}. The sky and background sources were subtracted as detailed in \cite{Tabatabaei_1_07}.
We also used the far-infrared (FIR) maps taken at 70$\mu$m and 160$\mu$m with the PACS instrument \citep{Poglitsch} onboard the Herschel space telescope as part of the Herschel M\,33 extended survey open time key project {\it HerM33es} \citep{Kramer10}. The PACS data were reduced using the scanamorphos algorithm \citep{Roussel_12} as discussed in detail in \cite{Boquien_11,Boquien_15}.
 The CO(2-1) line emission was observed with the IRAM-30m telescope as detailed in \cite{Gratier} and \cite{Druard}. The HI-21\,cm line was mapped with the VLA \citep{Gratier}.

{All maps used in our analysis were convolved} to the resolution of the 1.5\,GHz image ($\sim 15\arcsec$) and projected to the same grid and center position. We applied the dedicated convolution kernels provided by \cite{Aniano_11} for the Spitzer/Herschel maps and Gaussian kernels for the rest.

\begin{figure*}
	\begin{center}
		\resizebox{\hsize}{!}{\includegraphics*{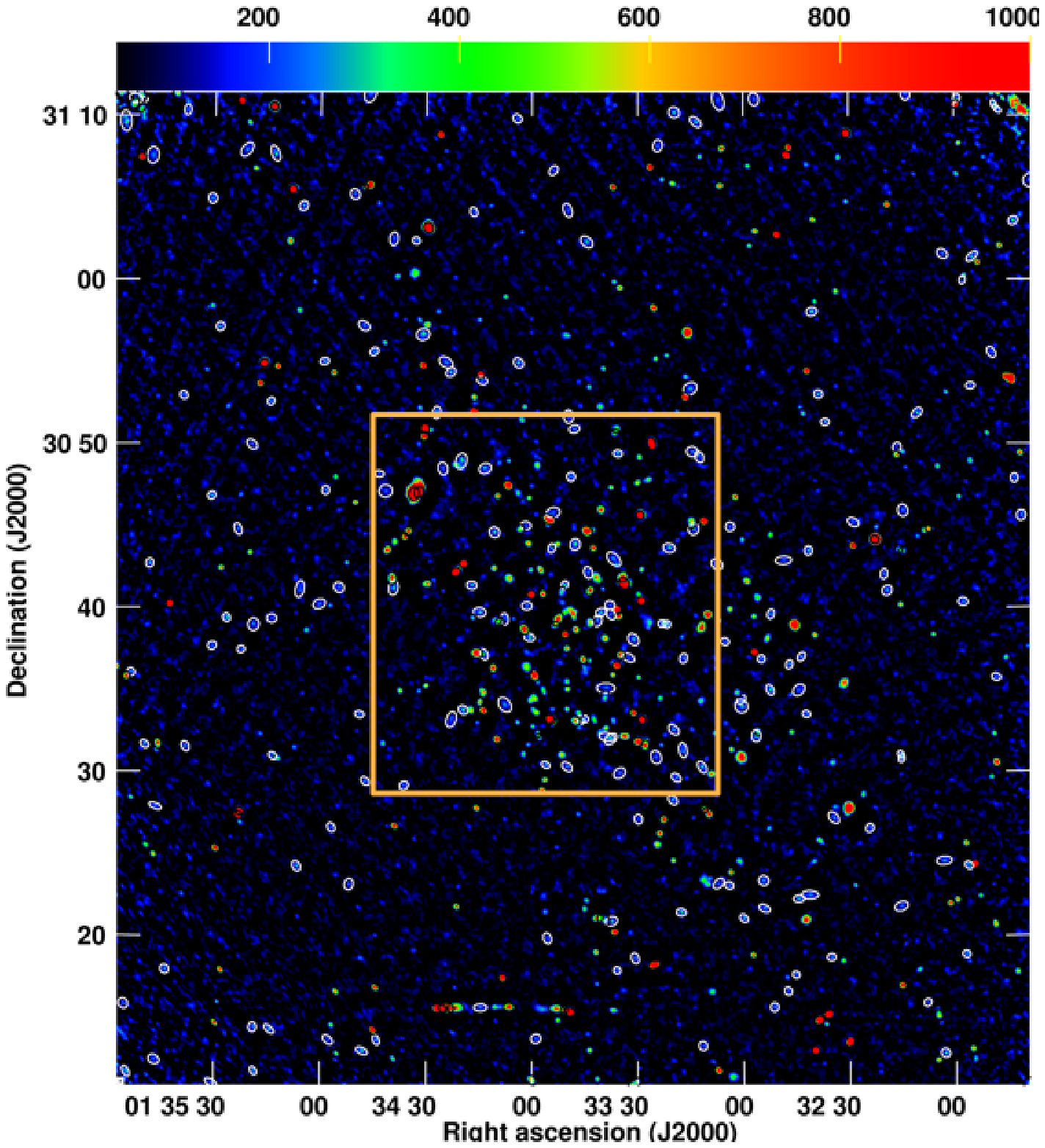}\includegraphics*{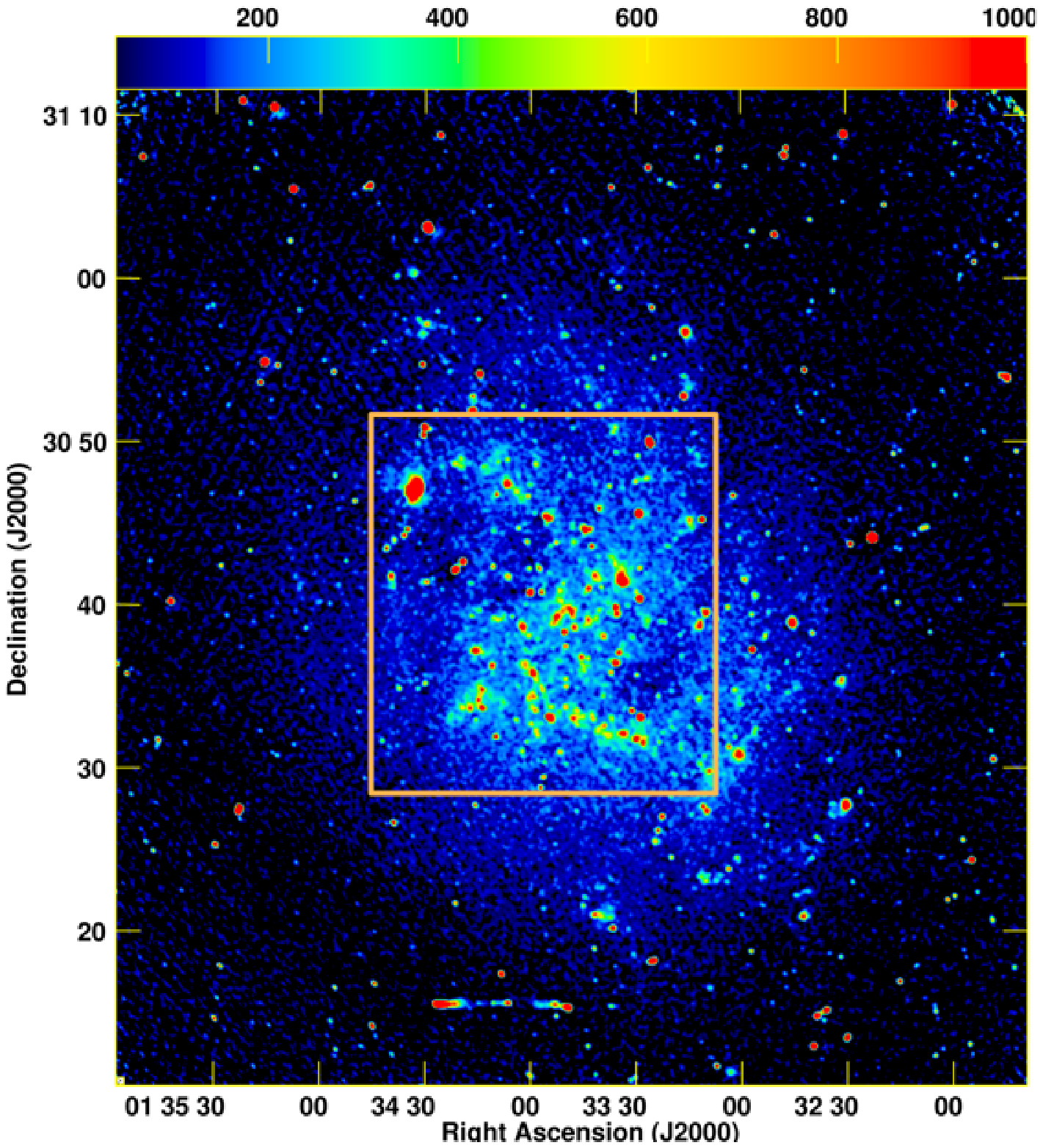}} 
		\resizebox{\hsize}{!}{\includegraphics*{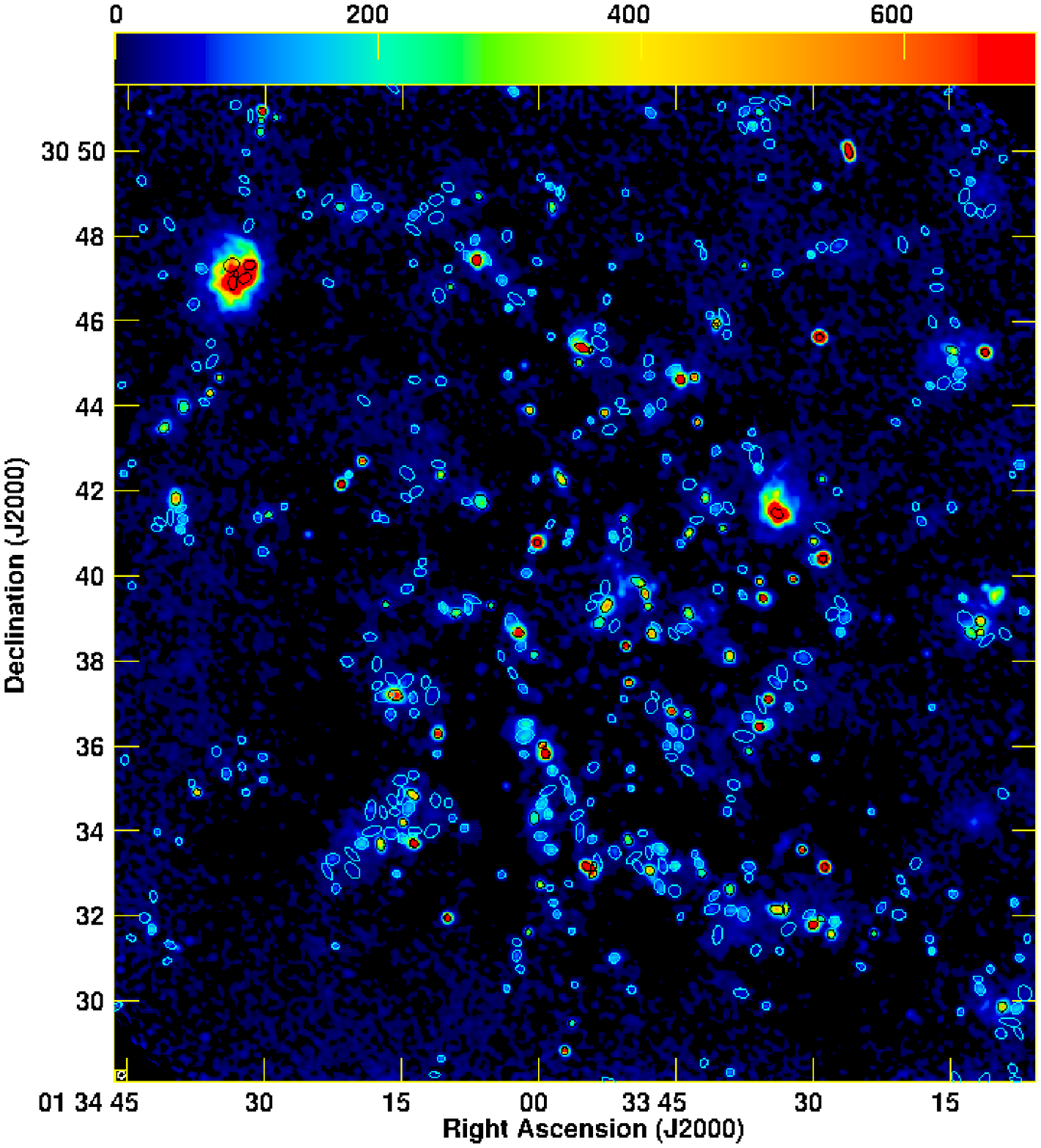}\includegraphics*{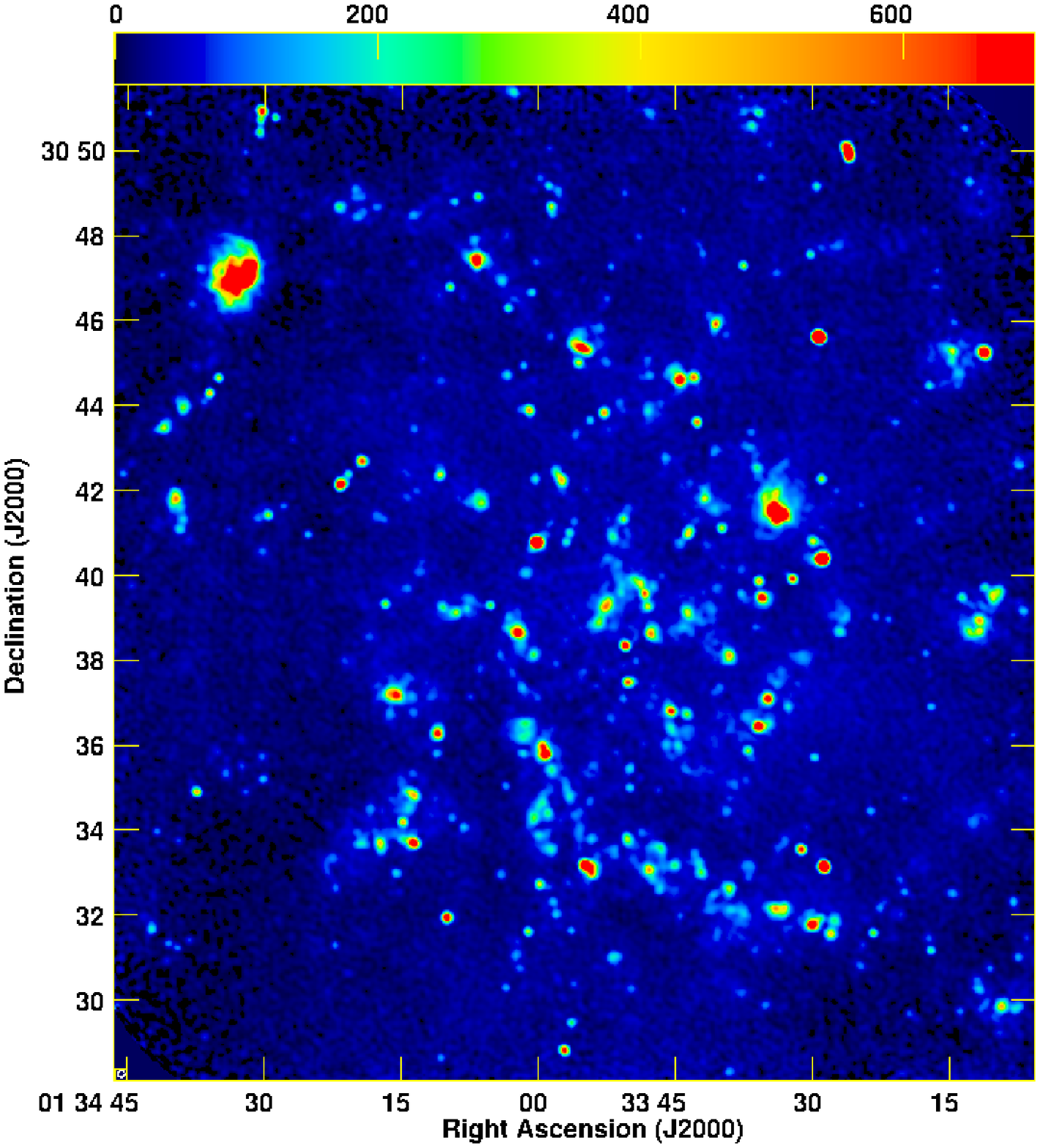}}
		\caption[]{Observed radio continuum emission at 1.5\,GHz ({\it top}) and  6.3\,GHz ({\it bottom}) before ({\it left}) and after correcting for missing short spacing information ({\it right}). The angular resolution of 15\arcsec~at 1.5\,GHz and 9\arcsec.35~at 6.3\,GHz are shown as circles in the bottom-left corner of each map. {In the left panels, the radio sources found at $>$5\,$\sigma$ rms are indicated by ellipses (different colours used for a better contrast against background). Angular size and position angle of the sources extracted (see Sect.~\ref{sec:pop}) are listed in Tables~\ref{tab:source1}~and~\ref{tab:source6}. Rectangles on top panels indicate the corresponding area observed at 6.3\,GHz shown in the bottom panels.}  }
		\label{fig:1.5}
	\end{center}
\end{figure*}

\subsection{Short Spacing Correction}

\label{sec:method} %
Interferometric observations miss much of the extended structures of the galaxy due to missing short spacings. The VLA observations are insensitive to angular scales larger than $\theta \sim \lambda/{\rm B_{min}}$ that is  $\sim\,5\arcmin$ and 22\arcmin\ at 6.3\,GHz (or 4.8\,cm) and 1.5\,GHz (or 20\,cm), respectively,  corresponding to the shortest baselines of ${\rm B_{min}}\sim$\,35\,m at C and D configurations. This results in negative bowl artifact around strong sources. For the short spacing correction (SSC), we combined the VLA maps with the Effelsberg single-dish data at corresponding frequencies presented in \cite{Tabatabaei_2_07} in the uv plane {using the AIPS task 'IMERG'. This program merges two input images by fourier transforming both, normalizing the second (the lower resolution map) to amplitudes within a uv annulus of the first, then producing an output transform plane consisting of the inner plane from the second input, a combination of the two within the annulus and the outer plane from the first. The output image is the back transform of this merged uv plane\footnote{see ${\rm http://www.aips.nrao.edu/cook.html}$}. Before running IMERG, the VLA and Effelsberg maps were projected into the same geometry and exact frequency \citep[using $\alpha=-0.7, S\propto \nu^{\alpha}$,][]{Tabatabaei_2_07} taking into account that the number of pixels in the map must be an integer power of 2 in both axes. The uv range (or min. and max. baselines in k$\lambda$) which defines the annular region of assumed overlap between the low and high resolution images should be set in this program so that the combined image has 1) an integrated flux density  and 2) a distribution of diffuse emission equal/similar to that of the low-resolution single-dish data. At 6.3\,GHz, these are achieved using a uv range of  0.739--1.236\,k$\lambda$. At 1.5\,GHz, a few strong sources are subtracted from the VLA and Effelsberg maps before merging them to avoid distortion effects caused by them. The uv range adopted to the residual maps is 0.175--0.236\,k$\lambda$. Those few sources are then re-added to the combined map. }
The rms noise in the resulting maps are about 12\,$\mu$Jy per 9\arcsec.35 at 6.3\,GHz and 48\,$\mu$Jy per 15\arcsec beam  at 1.5\,GHz.

After the SSC, the integrated flux density increases by about 50\% at 1.5\,GHz. We note that the negative pixels of the VLA map were put to zero before integrating its flux density, otherwise a much higher increase is obtained. At 6.3\,GHz, the SSC leads to $\simeq 60\%$ increase in the integrated flux density  in the inner $18\arcmin\times18\arcmin$ of the galaxy.

\begin{figure}
	\begin{center}
		\resizebox{8cm}{!}{\includegraphics*{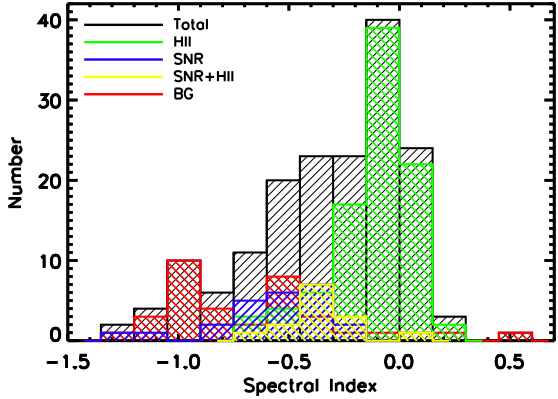}}
		\caption[]{Histogram of the spectral index of radio sources  in the inner $18\arcmin\times18\arcmin$  disk of M\,33 with $>5 \sigma$ detections at both 1.5\,GHz and 6.3\,GHz frequencies, that are L and C bands, respectively. }
		\label{fig:hist_so}
	\end{center}
\end{figure}

\section{Thermal and Nonthemal Separation Technique}
The radio maps at the frequencies observed here show a mixture of two main radiation mechanisms, the free-free emission from thermal electrons ${I^{\rm th}_{\nu}}$ and the nonthermal synchrotron emission from relativistic CR electrons (${I^{\rm th}_{\rm nt}}$, i.e., ${I^{\rm obs}_{\nu}} = {I^{\rm th}_{\nu}} + {I^{\rm nt}_{\nu}}$), referring to different origins and source populations such as HII regions, supernova remnants, background radio galaxies, and diffuse emission, {and to} the ionized and magnetized ISM. In order to dissect these sources and map the thermal and nonthermal ISM in M\,33, we adapt the Thermal Radio Template technique (TRT) developed by \cite{Tabatabaei_3_07}, in which the dust optical depth (proportional to dust mass and extinction) is first mapped across the galaxy.  Then the H$\alpha$ emission, as the most direct and physically motivated tracer of the free-free emission, is corrected for attenuation by dust. The nonthermal RC emission can be determined by subtracting from the observed RC the thermal radio emission calculated from the de-reddened H$\alpha$ emission. These steps are explained as follows.
\subsection{De-reddening the H$\alpha$ Emission}\label{dered}
To de-redden the H$\alpha$ emission, we first determine the amount of dust that is attenuating the H$\alpha$ photons using the {Herschel PACS} FIR data at 70\,$\mu$m and 160\,$\mu$m following \cite{Tabatabaei_3_07}. Assuming a modified black body radiation, the dust colour temperature $T$ and optical depth at 160\,$\mu$m, $\tau_{160}$ are obtained from the following equations,
\begin{equation}
\frac{I_{70}}{I_{160}}=\frac{\nu_{70}^{\beta}}{\nu_{160}^{\beta}} . \frac{B_{70}(T)}{B_{160}(T)},
\end{equation}

\begin{equation}
I_{160\mu{\rm m}} = B_{160}(T)\, [1 - e^{-\tau_{160}}],
\end{equation}
where $I$ denotes the observed intensity, $B(T)$ the Planck function, and $\nu$ the
frequency, and $\beta$ the dust emissivity index. In a detailed study of dust physical properties in M\,33, \cite{Taba_14} showed that $\beta$ changes across the galaxy from a mean value of $\beta\simeq\,1.8-2$ in the inner 4\,kpc radius of the galaxy to 1.5 and smaller beyond this radius. Hence, we adopted these variations of $\beta$ using the above equations.
The dust optical depth at 160\,$\mu$m is then converted to that at 6570\AA{}, i.e., the wavelength of the H$\alpha$ emission, using the dust extinction coefficient per unit mass at the corresponding wavelengths, $\tau_{\rm H_{\alpha}}= \kappa_{{\rm H}\alpha}/ \kappa_{160\mu {\rm m}} . \tau_{160} =\,2200 \, \tau_{160}$ \citep[e.g.][]{krugel}. {This interpolation uses a standard dust model and a Milky-Way extinction curve with the total/selective extinction ratio of $R_v=3.1$ \citep{Weingartner}. Studies of the dust extinction curves toward individual sight lines in M\,33 show that $R_v$ is similar to that of the Milky Way on average \citep{wang}.} The resulting $\tau_{\rm H_{\alpha}}$ varies from 0 to 1 across M\,33 with a mean value of $\simeq$ 0.4. Therefore, in agreement with our previous study at much lower resolution of 90\arcsec \citep{Tabatabaei_3_07}, M\,33 is almost transparent for photons with $\lambda \simeq 6560$\AA\ propagating towards us.  For an inhomogenous distribution of dust and ionized gas, only about one third of the dust present is effectively attenuating the H$\alpha$ photons \citep{Dickinson} and hence the emission received is given by $I = I_0 \,\,e^{-\tau_{\rm eff}}$ with $\tau_{\rm eff} = \frac{1}{3}. \tau_{{\rm H}\alpha}$.  This means that, on average, only about 14\% of the H$\alpha$ emission is absorbed by dust in M\,33.

{Using the full Herschel data set, we presented maps of dust mass and temperature in M\,33 by modeling the dust SED \citep{Taba_14} at about 40\arcsec~angular resolution (resolution of the SPIRE 500\,$\mu$m data). These products are not used here because their resolution is poorer than the resolutions of the RC maps. We, however, note that the difference in the resulting dust optical depths obtained based on the two methods is less than 10\% at 40\arcsec~angular resolution.
	
A more straightforward de-reddening method of the H$\alpha$ emission is the Balmer-decrement-ratio technique, as it does not depend on the dust filling factor along the line of sight and directly measures the attenuation of photons emerging from ionized gas \citep[see e.g.,][]{calzetti_0,Kreckel,Kouroum,Tacchella_22}. We used this method to trace the thermal free-free emission in the center of NGC1097 for the first time \citep{Taba_18}. For M\,33, this method is not used because the only map of the Balmer line with reliable calibration available is the H$\alpha$ map. Using the MUSE observations of the H$\alpha$ and H$\beta$ lines in the Magellanic Clouds, \cite{Hassani} performed a detailed comparison of the Balmer-decrement-ratio technique with the dust optical depth method explained above. They mapped the fraction of total dust content attenuating the H$\alpha$ emission in these galaxies and showed that on average the two methods agree well on scales of GMCs and larger. }

\subsection{Mapping the RC Components}

The radio free-free intensity in mJy/beam, $I^{th}_{\nu}$, is related to the brightness temperature $T_b$ in Kelvin through the Rayleigh Jeans relation:
\begin{equation}
I^{th}_{\nu} = \frac{\theta_i \theta_j}{1.36 \lambda^2} T_b,
\end{equation}
with $\theta_i$ and $\theta_j$ the beam width along the major and minor axes in arcsec and $\lambda$ the radio emission wavelength in cm. In an ionized gas, $T_b$ is related to the H$\alpha$ intensity, $I_{{\rm H}\alpha}$, in erg\,cm$^{-2}$\,s$^{-1}$ sr$^{-1}$ via
\begin{equation}
\left \{ \begin{array}{ll}
{T_b=T_e(1-e^{-f(T_e)\,I_{{\rm H}\alpha}})} ,  \\
f(T_e)=3.763\,a\,\nu_{\rm GHz}^{-2.1}\, T_{e4}^{-0.3}\, 10^{\frac{0.029}{T_{e4}}},   %
\end{array} \right.
\end{equation}
where $T_e$ is the electron temperature in K, $T_{e4}$ is $T_e$ in units of $10^4$\,K, and taking into account the contribution from singly ionized He. {This relation uses the Gaunt factor approximation of \cite{Altenhoff} with $a\simeq 1$.} In case the ionized gas is optically thin for thermal radio continuum photons, the following relation holds between the intensities of the thermal radio and H$\alpha$ emission
\begin{eqnarray}
\left(\frac{I^{\rm th}_{\nu}}{\rm mJ/beam}\right)&=&\left. 3.074 \times 10^{-35}\,\theta_i \theta_j\, \nu_{\rm GHz}^{-0.1}\,T_{e4}^{0.668}\right. \\
&& \left. \times \, 10^{\frac{0.029}{T_{e4}}} \left(\frac{I_{{\rm H}\alpha}}{\rm erg\,cm^{-2}\,s^{-1} sr^{-1}}\right) \right.. \nonumber
\end{eqnarray}
{The present study is however based on the more general condition given by Eq.~(4) adopting an electron temperature of $T_e=10^4$\,K. The resulting thermal fraction ($I^{\rm th}_{\nu}/I^{\rm obs}_{\nu}$) drops by about 20\% assuming $T_e=7000$\,K \citep{Taba_13}. As shown by \cite{Tabatabaei_3_07}, the thermal fraction is by at least a factor of 2 less sensitive to variations in $T_e$ than to variations in the synchrotron spectral index ($\alpha_n$, $I_{\rm nt}\propto \nu^{\alpha_n}$), assumed to be fixed in the classical separation technique. After mapping the thermal intensity, the} nonthermal intensity, ${I^{\rm nt}_{\nu}} = {I^{\rm obs}_{\nu}} - {I^{\rm th}_{\nu}}$, is calculated for each pixel of the maps at 1.5 and 6.3 GHz. {We note that local variations in the electron temperature and/or the NII-to-H$\alpha$ ratio can lead to an over-prediction of the thermal emission and a negative nonthermal intensity as occurs in some parts of bright HII regions in M33 (Fig.~\ref{fig:TNT20}). }

\section{Results}
\label{sec:result}
\subsection{Overall Distribution of RC}
The maps shown in Fig.~\ref{fig:1.5} illustrate structures emitting RC across the entire disk of M\,33 at 1.5\,GHz and the inner $18\arcmin \times 18\arcmin$ area  at 6.3\,GHz. While the structures are limited to radio sources before the SSC (left panels), clumps surrounding the bright sources, spiral arms, extended structures, and diffuse emission from the disk are evident only after the SSC (right panels). We now investigate the nature and origin of each RC component.

\begin{figure*}
	\begin{center}
		\resizebox{\hsize}{!}{\includegraphics*{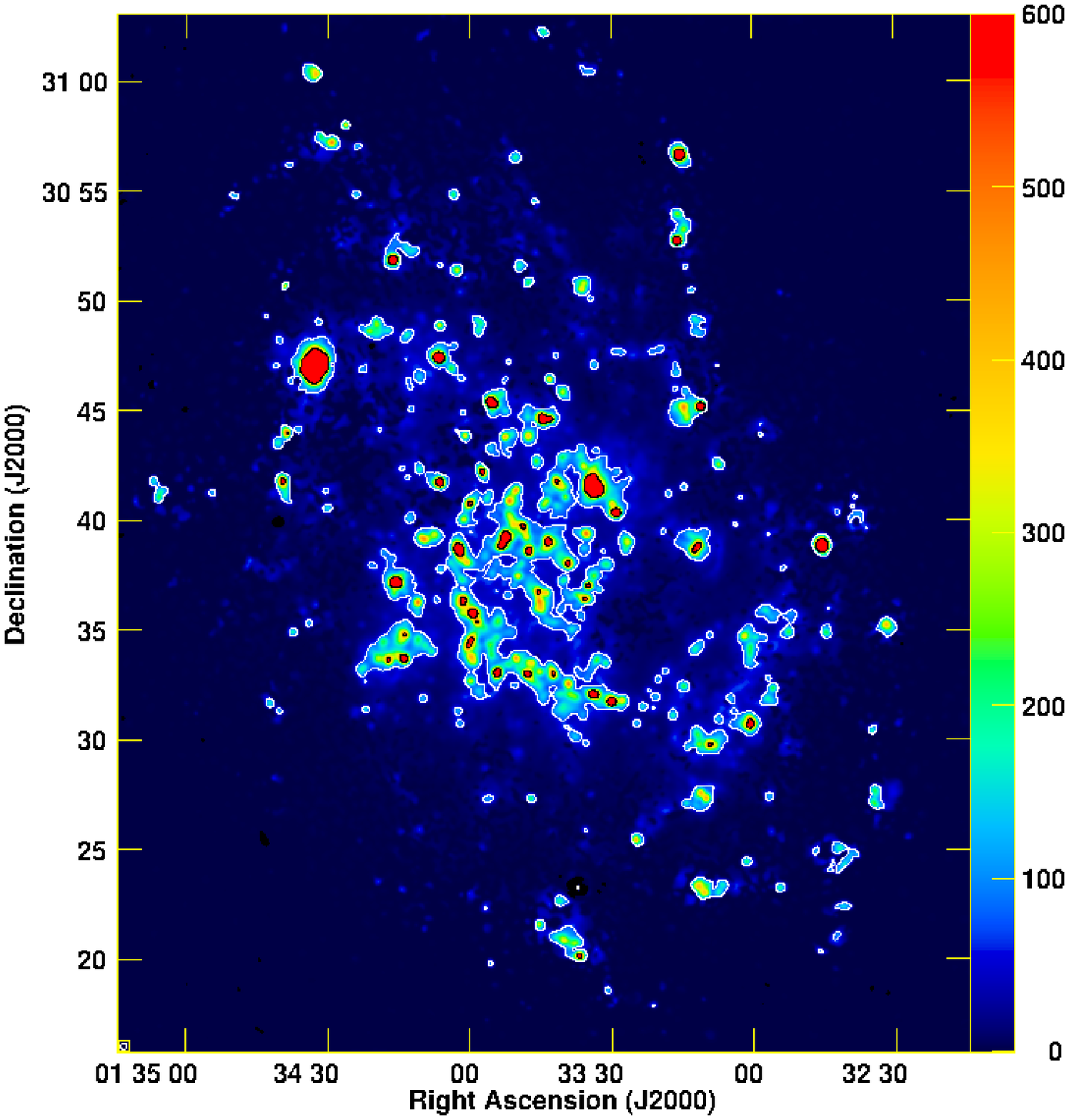}\includegraphics*{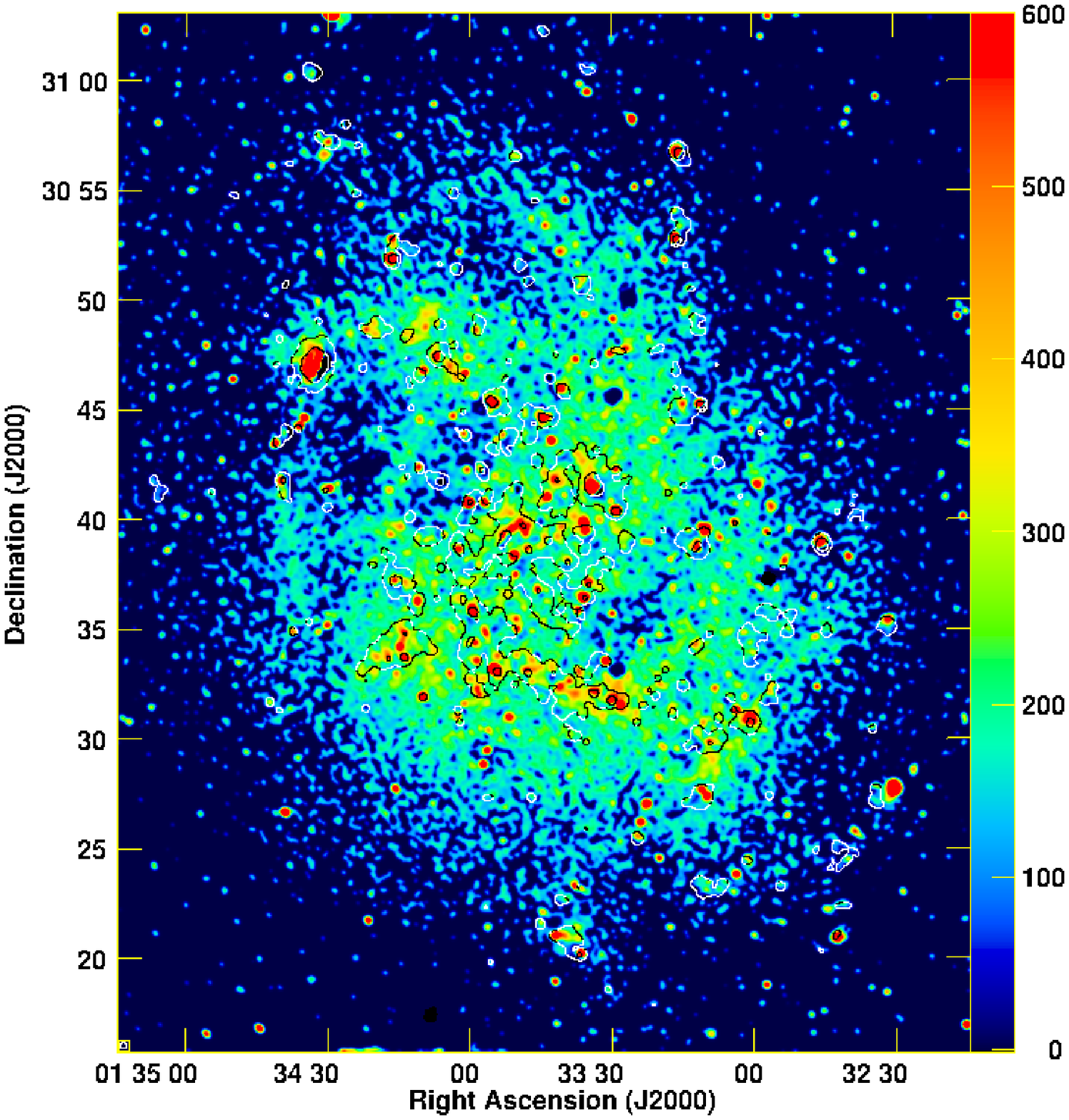}}
		\resizebox{\hsize}{!}{\includegraphics*{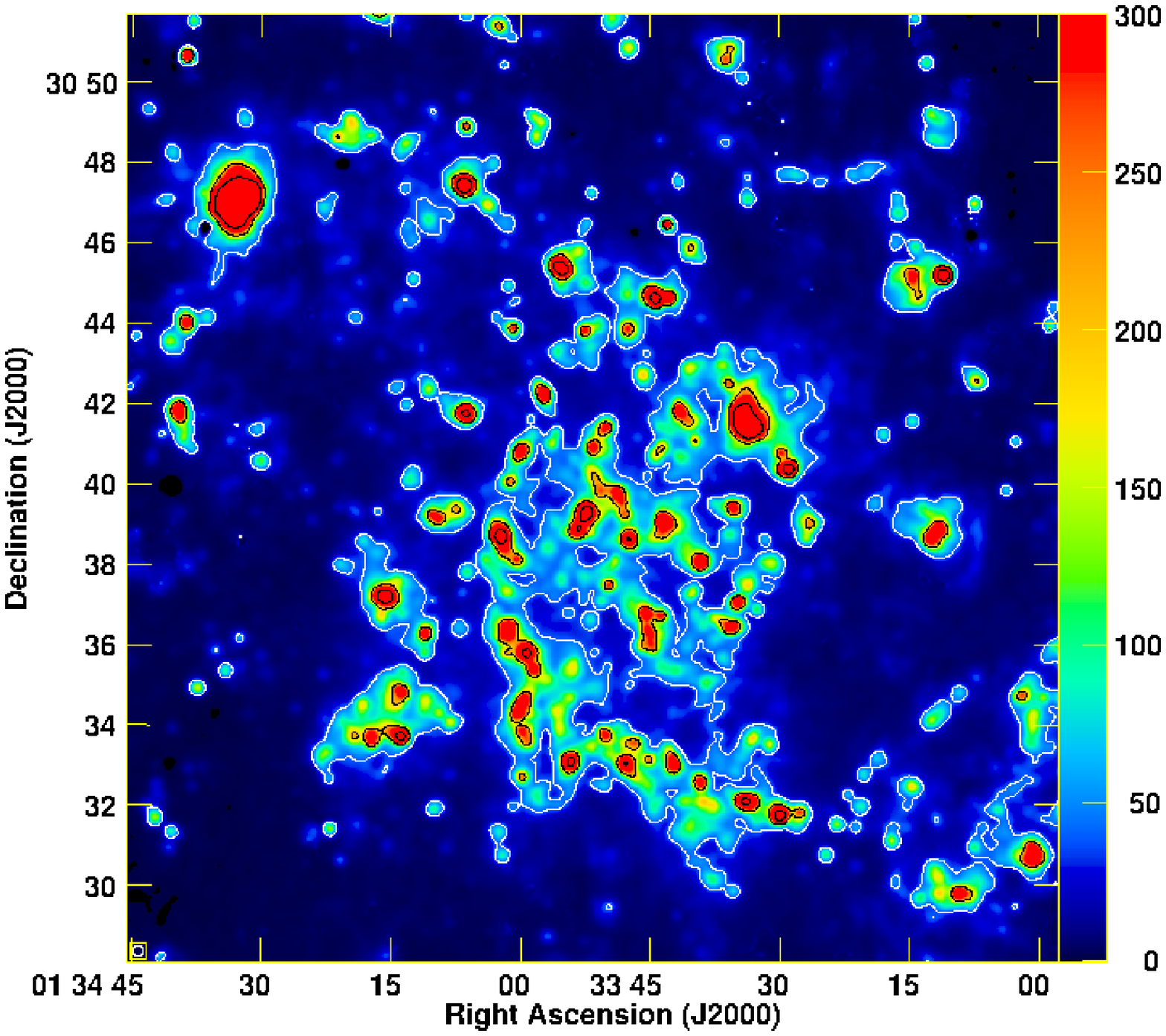}\includegraphics*{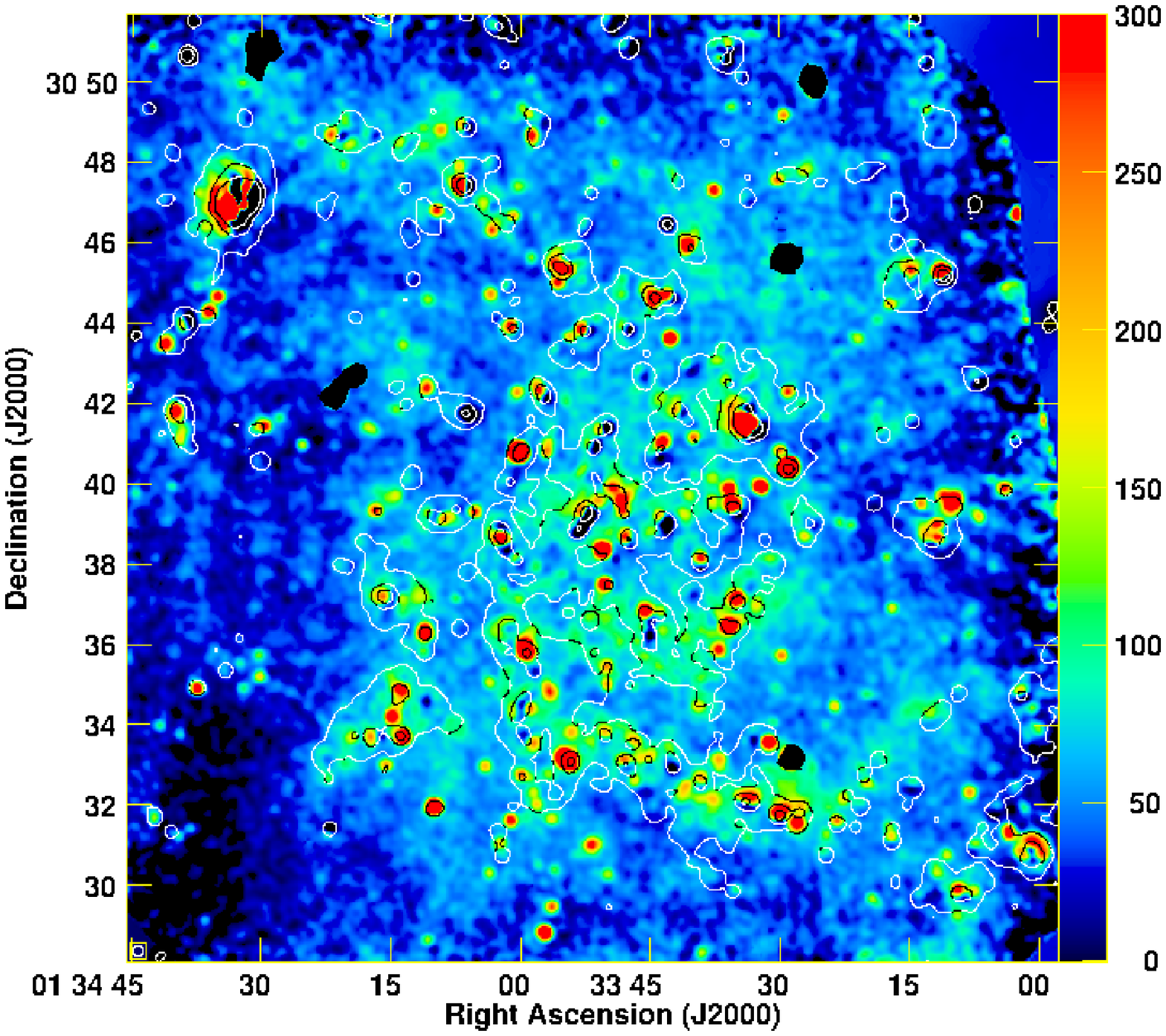}}
		\caption[]{ {\it Top:} the thermal ({\it left}) and nonthermal ({\it right}) RC emission in $\mu$Jy/beam at 1.5\,GHz overlaid with contours of the thermal free-free emission at levels 70, 500\,$\mu$Jy/beam. {\it Bottom:} same as above at 6.3\,GHz overlaid with the thermal emission contours at levels 40,200, 600\,$\mu$Jy per/beam. The maps at 6.3\,GHz cover only the inner $18\arcmin \times 18\arcmin$ galaxy. The half power beam width of 15\arcsec~is shown by a circle in the left corner of each image. Background radio sources with peak intensities of $\geq 4.6$\,mJy/beam at 1.5\,GHz ($\geq 1$\,mJy/beam at 6.3\,GHz) are subtracted from the nonthermal maps.  }
		\label{fig:TNT20}
	\end{center}
\end{figure*}

\begin{table*}
	\begin{center}
		\caption{Radio sources detected in M\,33 at 1.5\,GHz (the full table is available as supplementary material). Parameters Maj, Min, and PA refer to the deconvolution of the best fit major axis, minor axis, and position angle, respectively. }
		\begin{tabular}{ l l l l l l l l }
			\hline
			ID &    Peak flux density &   Integrated flux density&    RA &     DEC &   Maj &    Min &   PA   \\
			& \,\,\,$\mu$Jy/beam&\,\,\,\,\,\,\,\,\,\,\,\, \,\,\,\,\,\,$\mu$Jy&     ($^h$ $^m$ $^s$)           &       ($^{\circ}$ $\arcmin$ $\arcsec$)      &  arcsec  & arcsec  &deg. \\
			\hline
			\hline
			1 &    183 $\pm$41 &   558$\pm$ 162& 01 31 38.08 & 31 06 00.85 & 27.4 $\pm$6.2&   25.1 $\pm$5.7 &  110 $\pm$105\\
			2&   249 $\pm$44&  290 $\pm$ 85 &01 31 39.08 &  30 11 56.87 & 20.5 $\pm$3.7 & 12.8 $\pm$2.3 & 177 $\pm$15 \\
			3 &   368 $\pm$ 41&  1028 $\pm$ 152& 01 31 39.28 & 30 10 49.94&  12.9 $\pm$0.9&  10.7 $\pm$0.8 &  41 $\pm$16\\
			4 &   1081 $\pm$ 42&  2610 $\pm$ 140& 01 31 39.70 &  31 10 09.06&  30.2 $\pm$1.2 & 18.0 $\pm$0.7 & 11 $\pm$3\\
			5 &  1319 $\pm$ 43&  2400 $\pm$ 110& 01 31 41.13&  31 10 21.43&  22.7 $\pm$0.7 & 18.0 $\pm$0.6 & 154 $\pm$ 6 \\
			6 &  249 $\pm$ 43&  354 $\pm$ 96& 01 31 41.15&  30 45 35.83&  19.3 $\pm$3.4&   16.5 $\pm$2.9&  169 $\pm$46 \\
			7 &   933 $\pm$45& 1050 $\pm$ 84 & 01 31 41.48&  30 49 23.76&  16.6 $\pm$0.8& 15.3 $\pm$0.7 & 54 $\pm$23\\
			\hline
			\hline
		\end{tabular}
		\label{tab:source1}
	\end{center}
\end{table*}

\begin{table*}
	\begin{center}
		\caption{Radio sources detected in the inner $18\arcmin \times 18\arcmin$ M\,33 at 6.3\,GHz (the full table is available as supplementary material). Parameters Maj, Min, and PA refer to the deconvolution of the best fit major axis, minor axis, and position angle, respectively. }
		\begin{tabular}{ l l l l l l l l }
			\hline
			ID &    Peak flux density &   Integrated flux density&    RA &     DEC &   Maj &    Min &   PA   \\
			& \,\,\,$\mu$Jy/beam&\,\,\,\,\,\,\,\,\,\,\,\, \,\,\,\,\,\,$\mu$Jy&     ($^h$ $^m$ $^s$)           &       ($^{\circ}$ $\arcmin$ $\arcsec$)      &  arcsec  & arcsec  &deg. \\
			\hline
			\hline
			1 &    62.7 $\pm$10.8 &   78.20$\pm$ 21.7& 01 33 06.64 & 30 31 28.17 & 12.3 $\pm$2.1&   8.8 $\pm$1.5 &  4 $\pm$21\\
			2&   37.2 $\pm$9.9&  183.1 $\pm$ 58.0 & 01 33 06.89&  30 30 11.02 & 22.4 $\pm$5.9 & 19.3 $\pm$5.1 & 29 $\pm$72 \\
			3 &   142.7 $\pm$ 10.5&  224.97 $\pm$ 25.07& 01 33 06.95 & 30 10 10.42&  12.9 $\pm$0.9&  10.7 $\pm$0.8 &  41 $\pm$16\\
			4 &   134.6 $\pm$ 10.3&  293.6 $\pm$ 31.1& 01 33 07.38 &  30 42 37.12&  14.7 $\pm$1.1 & 12.97 $\pm$0.99 & 79 $\pm$24\\
			5 &  124.3 $\pm$ 10.7&  169.1 $\pm$ 22.9& 01 33 07.48&  30 31 00.96&  11.4 $\pm$1.0 & 10.4 $\pm$0.9 & 87 $\pm$ 38 \\
			6 &  41.4 $\pm$ 10.3&  85.4 $\pm$ 29.9& 01 33 07.59&  30 29 14.87&  20.1 $\pm$5.0&   8.9 $\pm$2.2&  31 $\pm$11 \\
			7 &   171.2 $\pm$10.1& 526.9 $\pm$ 40.0 & 01 33 07.76&  30 29 49.41&  18.3 $\pm$1.1&  14.7 $\pm$0.9& 128 $\pm$11\\
			\hline
			\hline
		\end{tabular}
		\label{tab:source6}
	\end{center}
\end{table*}

\subsubsection{Radio source population}
\label{sec:pop}
Before the SSC, we extracted radio sources with scales of giant HII regions and smaller, i.e.,  $<200$\,pc ($57\arcsec$), from the observed mosaics at 6.3 and 1.5\,GHz using the AIPS source extraction task called {\it SAD}.
The task finds potential islands of emission down to a threshold level in a few rounds of searching. Then each island is fitted by one or multiple elliptical Gaussians (in case of multiple peaks) while allowing the fit to find angular sizes of sources starting with the beam size.

Setting the threshold to 5\,$\sigma$ rms,  516 sources were found with peak intensities $>$ 200\,$\mu$Jy per 15\arcsec ~beam at 1.5\,GHz (Fig.~\ref{fig:1.5}, top-left).
At 6.3\,GHz, we found 488 sources with a threshold of 35\,$\mu$Jy per 9.35\arcsec (5\,$\sigma$ rms) in the inner $18\arcmin \times 18\arcmin$ (Fig.~\ref{fig:1.5}, bottom-left)\footnote{The catalogues at original resolutions and map sizes are presented as an Appendix in the electronic version.}.  For sources having the beam size or smaller, the peak intensity is taken as the integrated flux density. We note that the number of radio sources found at 6.3\,GHz is by a factor of 2.6 larger than that found by \cite{Gordon_s_99} within 20\arcmin\ of the center of M\,33.

As the full spectrum analysis is limited to the C-band observations' field of view, we build a subsample of sources in the inner $18\arcmin \times 18\arcmin$ area of M\,33. To obtain their 1.5--6.3\,GHz spectral index, the 6.3\,GHz map is first convolved to the same resolution as the 1.5\,GHz map. At this resolution, about 343 sources were fitted at 6.3\,GHz with a threshold of 50\,$\mu$Jy (5\,$\sigma$ rms), while only about 204 were found in the same area at 1.5\,GHz due to this image's lower sensitivity. {The results are cross-matched by considering a maximum source separation of half of the L-band synthesized beam. About} 167 sources are detected emitting at both 1.5\,GHz and 6.3\,GHz for which we determined the spectral index $\alpha$ ($S\sim \nu^{\alpha}$). We note that it is, of course, possible to build a larger sample by measuring the 1.5\,GHz integrated flux densities at the source positions of the more sensitive 6.3\,GHz observation. {However, to measure the spectral index more robustly, a consistent S/N at different frequencies is required that is taken to be 5\,$\sigma$ rms.} Table~A1 lists the source properties in the inner $18\arcmin \times 18\arcmin$ of M\,33.

To identify these sources, we first cross-correlated the final sample with catalogs presented by \cite{Gordon_s_99}, \cite{Long_10}, and \cite{White}. {These authors classified radio sources mainly using their [SII]:H$\alpha$ ratios and, when available, characteristics of their X-ray emission \citep{Long_10,White}. Generally, SNRs have higher [SII]:H$\alpha$ ratios (0.4 to 1) than HII regions because their radiative shocks contain cooling zones where [SII] is the dominant ionization  state  of  S,  whereas  in (bright) HII regions [SIII] is the dominant ionization  state \citep{Dodorico,Levenson}. } From our selected sample of sources, about 51\%  are {\rm found to be} known HII regions (three of which emit X-ray as well), 15\%  SNRs, 8\%  HII+SNRs, 15\%  BG sources. The remaining radio sources have no classified counterparts but some coincide with X-ray sources. We determine a mean {radio} spectral index of  $\alpha=-0.12 \pm 0.16$ for the HII regions (where the error is the standard deviation) a value that agrees with an optically thin thermal spectrum ($\alpha \simeq\,-0.1$). The spectral index is steeper for the combined HII+SNR sources with a mean $\alpha= -0.37 \pm 0.16$. For sources identified as SNRs, we obtain $\alpha= -0.60 \pm 0.30$ and for BGs,~$\alpha= -0.65 \pm 0.44$. 

We further investigate if the sources classified as unknown in the \cite{White} catalogue belong to M\,33 or are BG sources. A robust determination of spectral index can help to identify different types of radio sources. As mentioned above, the emission is predominantly free-free in HII regions with a flat spectrum $\alpha \simeq -0.1$. In SNRs, the emission mechanism is synchrotron and the spectral index should be steeper than $-0.2$. Emission from background radio galaxies is also synchrotron with a spectral index that is often steeper than $-0.5$. It can, however, be flatter and even inverted for very compact radio sources such as some quasars and other active galactic nuclei (AGNs) due to synchrotron self-absorption \citep[e.g.,][]{Kellermann}. Cross-correlation with H$\alpha$ emission can provide additional constraints for the internal sources as the H$\alpha$ line emission from BG radio sources is unlikely to fall in the velocity window adopted for M\,33. From the 18 unknown sources that have spectral index information, seven sources coincide with H$\alpha$ sources.
Their relatively flat spectral index with a mean $\alpha=-0.29 \pm 0.30$ makes them candidates of HII regions (or HII+SNRs) in M\,33. Eleven sources are taken as external to M\,33, increasing the total number of BGs to 36. We note that this addition does not change the BGs' mean $\alpha$ ($=-0.65 \pm 0.44$). Figure~\ref{fig:hist_so} shows a histogram of the spectral index $\alpha$ of the different source types.

\begin{figure*}
	\begin{center}
		\resizebox{\hsize}{!}{\includegraphics*{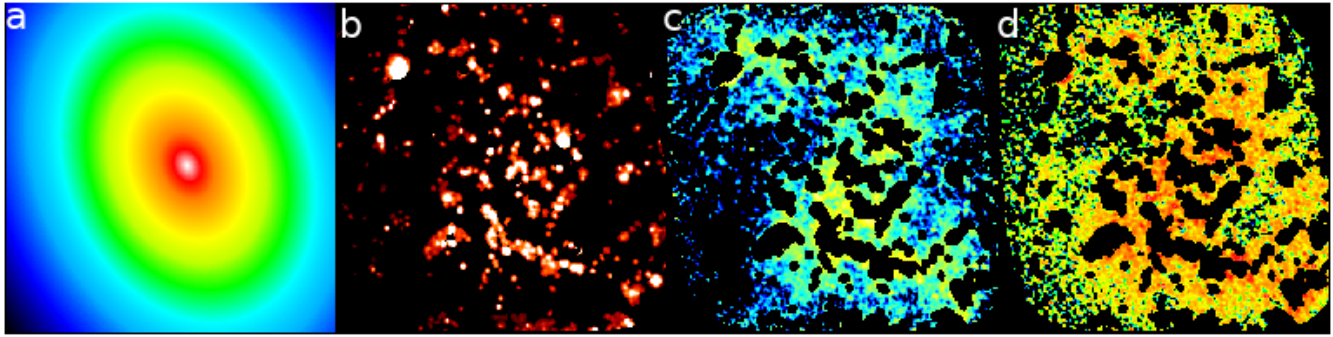}} 
		\caption[]{Structural decomposition of the RC emission from the inner $18\arcmin \times 18\arcmin$ part of M\,33: {\it a-} diffuse disk, {\it b-} sources, and {\it c-} extended structure at 6.3\,GHz. The field of view is the same as Fig.~\ref{fig:TNT20}-bottom panels.  Strong BGs are subtracted hence the sources indicate mostly the star-forming regions. Also shown is the extended emission at 1.5\,GHz from the same area as of the 6.3\,GHz extended emission ({\it d}). A quantitative comparison of the relative contribution of each structural component at different frequencies is shown in Table~\ref{tab:diffuse}. }
		\label{fig:disk}
	\end{center}
\end{figure*}

\subsubsection{Diffuse disk and extended RC emission}
\label{sec:struc}
The extracted sources  account for 34\% of the 1.5\,GHz emission and 44\% of the 6.3\,GHz emission from the inner $18\arcmin \times 18\arcmin$ of M\,33. {The remaining emission emerges from patches of extended structures (on scales of $<2$\,kpc) surrounding star-forming regions in spiral arms (photo-ionized or shocked-ionized regions) or from structures in between the arms (likely of nonthermal origin, see Sect.~\ref{sec:tnst}). On larger scales, the general diffuse disk of the galaxy can contribute significantly}.
Hence, to determine the relative contribution of the extended structures, the contribution from the diffuse disk must be subtracted. The surface brightness of spiral galaxies {generally} falls exponentially with galactocentric radius $R$ with an exponential scale length $l$. For the full disk of M\,33, the scale lengths of the RC emission are presented by \cite{Tabatabaei_2_07} at different frequencies\footnote{Exponential fits to the new VLA data lead to similar scale lengths at corresponding galactocentric radii. }.
The sky-plane projection of the diffuse RC disk is described as $I_{\rm d}(R)= {A\, exp(-R/l)}$ with $R$ the galactocentric radius.  After removing the point sources, we fit this model to only diffuse regions in interarm regions and/or outer parts to avoid contributions from other extended structures. The model parameters $A$ and $l$  are obtained using the  bisector least square fits \citep{Isobe} at different frequencies (Table~\ref{tab:diffuse}).
\begin{table*}
\begin{center}
\caption{Contribution of diffuse disk, extended emission, and discrete sources to the total integrated flux density in the inner $18\arcmin \times 18\arcmin$ area of M\,33 for the observed RC and its thermal and nonthermal components. The five birgthest background sources are excluded.   The parameters $a$ and $b$ are the diffuse disk model fit parameters  in $Y= b\,X - A$ with $Y= ln (I_{d})$, $X=R$, $A=ln (a)$, and $b=1/l$. }
\begin{tabular}{ l l l l l l l l}
\hline
$\nu$ & Component & $S_{\rm total}$& ${S_{\rm disk}}/{S_{\rm total}}$ &${S_{\rm extended}}/{S_{\rm total}}$ & ${S_{\rm source}}/{S_{\rm total}}$ &$a$ \, & $b$\\
  (GHz)  &  &  (mJy)   &  &  & &   ($\mu$Jy/beam)     & (kpc)$^{-1}$\\
\hline
\hline
1.5 & Observed & 812\,$\pm$\,68 &0.41\,$\pm$\,0.03 & 0.23\,$\pm$\,0.02 & 0.36\,$\pm$\,0.02 & 118$\pm$\,3 & 0.18$\pm$\,0.02\\
       & Nonthermal &610\,$\pm$\,60    &0.44\,$\pm$\,0.03 & 0.26\,$\pm$\,0.02  & 0.30\,$\pm$\,0.02 & 101\,$\pm$\,4 & 0.16\,$\pm$\,0.02\\
       & Thermal & 202\,$\pm$\,10  & 0.30\,$\pm$\,0.02 & 0.15\,$\pm$\,0.01  & 0.55\,$\pm$\,0.02 & 24\,$\pm$\,3 & 0.49$\pm$\,0.06\\
\hline
6.3  & Observed & 432\,$\pm$\,29 & 0.36\,$\pm$\,0.03 & 0.18\,$\pm$\,0.02 & 0.46\,$\pm$\,0.02 & 71\,$\pm$\,10 & 0.32$\pm$\,0.09\\
       & Nonthermal & 257\,$\pm$\,19 & 0.40\,$\pm$\,0.03 & 0.20\,$\pm$\,0.02 & 0.40\,$\pm$\,0.02  & 51\,$\pm$\,2 &0.27\,$\pm$\,0.01\\
       & Thermal & 175\,$\pm$\,8 & 0.30\,$\pm$\,0.02  & 0.15\,$\pm$\,0.01  & 0.55\,$\pm$\,0.02  & 21\,$\pm$\,3 & 0.49$\pm$\,0.06\\
\hline
\end{tabular}
\label{tab:diffuse}
\end{center}
\end{table*}

Subtracting the diffuse disk, we then separately map the extended structures. An example of this structural decomposition is shown in Fig.~\ref{fig:disk} for the RC emission at 6.3\,GHz. The integrated flux density and contribution of diffuse disk and extended structures in the inner part of M\,33 are then determined (Table~\ref{tab:diffuse}). The integrations are performed in rings in the plane of the galaxy around the centre out to a radius of 12\arcmin.75 after subtracting the background radio sources.

{The uncertainties in the integrated flux densities are determined taking into account the map fluctuations ($\delta_{\rm rms}=  \sigma_{\rm rms}\,\, \frac{a}{\theta} \, \sqrt{\frac{\rm N}{1.133}},$ with $\sigma_{\rm rms}$ is the rms noise level, $\theta$ the angular resolution, N the number of pixels, and $a$ the pixel size), calibration uncertainty ($\delta_{\rm cal}\simeq$\,3 per cent for the VLA, and 2 and 5 per cent at 1.5 and 6.3\, GHz, respectively, for the Effelsberg observations), and the uncertainty of the baselevel in the single-dish map ($\delta_{\rm base}~=~\sigma_{0}\,\, {\rm N_{\rm beam}}$ with $\sigma_{0}$ the baselevel uncertainty of the Effelsberg map $\sigma_{0} \simeq 0.2\,\, \sigma_{\rm rms}$). Hence, the total uncertainty is  given by  $\delta= \sqrt{\delta_{\rm cal}^2 + \delta_{\rm rms}^2 + \delta_{\rm base}^2}$. The errors in the integrated flux density of the maps are then propagated to obtain uncertainties in the ratios reported in Table~\ref{tab:diffuse}.  }

{At 6.3\,GHz, most of the RC emission emerges from unresolved sources (46\%$\pm$2\%), while the diffuse disk dominates at 1.5\,GHz  (41\%$\pm$3\%). The contribution of the extended structures to the total RC is about 23\% and 18\% at 1.5\,GHz and 6.3\,GHz, respectively.   }

\begin{figure}
	\begin{center}
		\resizebox{\hsize}{!}{\includegraphics*{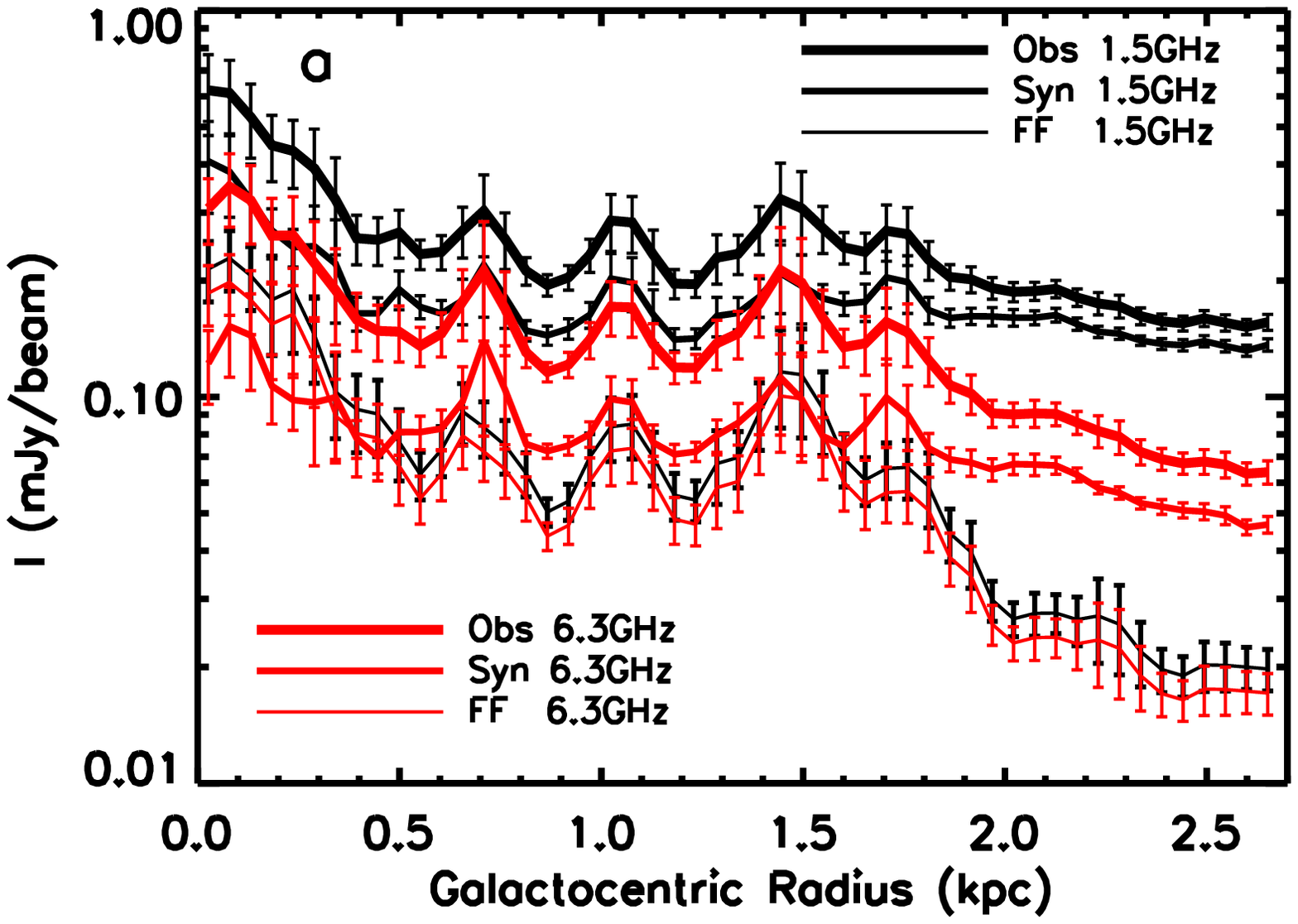}}
		\resizebox{\hsize}{!}{\includegraphics*{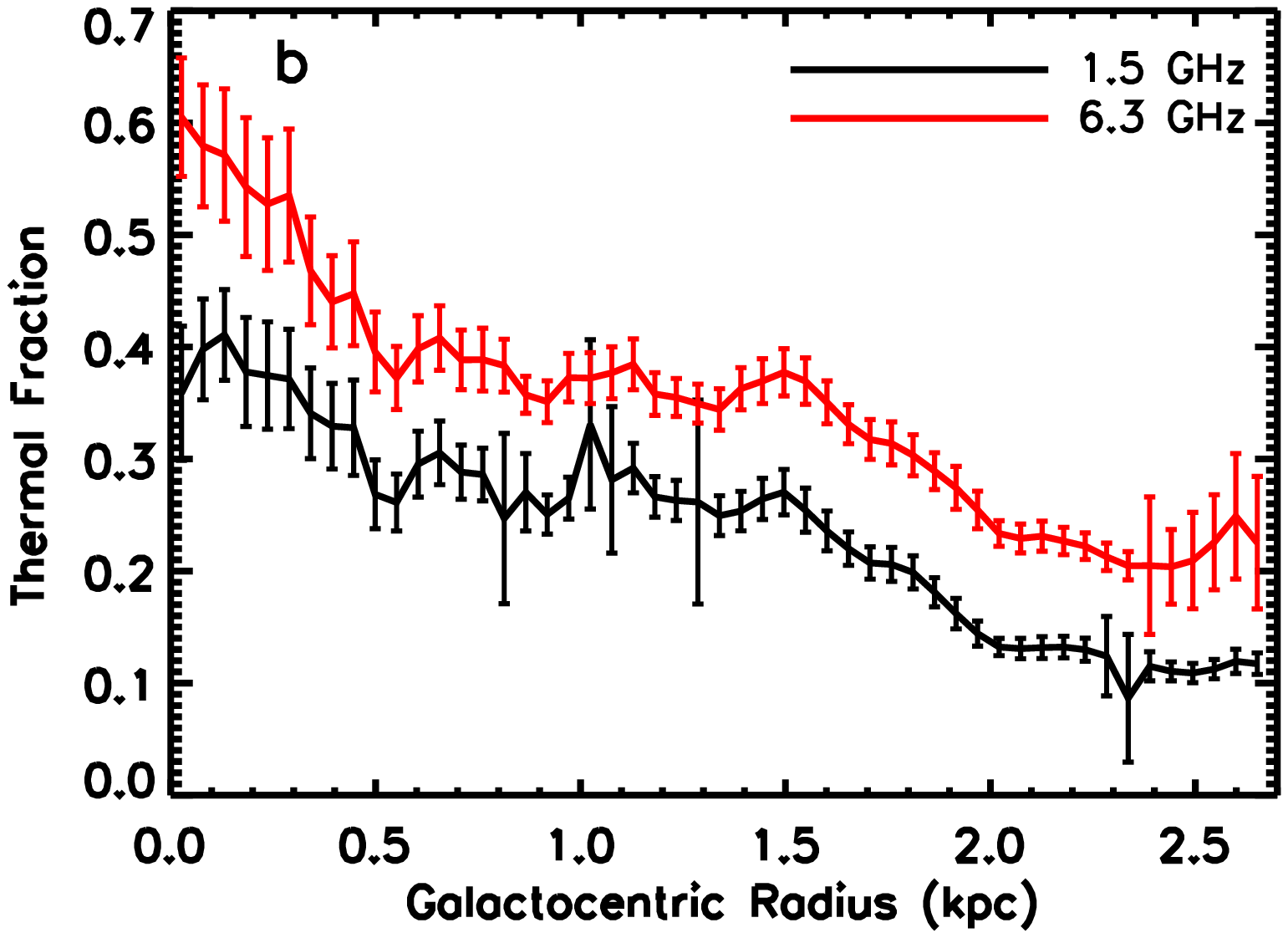}}
		\resizebox{\hsize}{!}{\includegraphics*{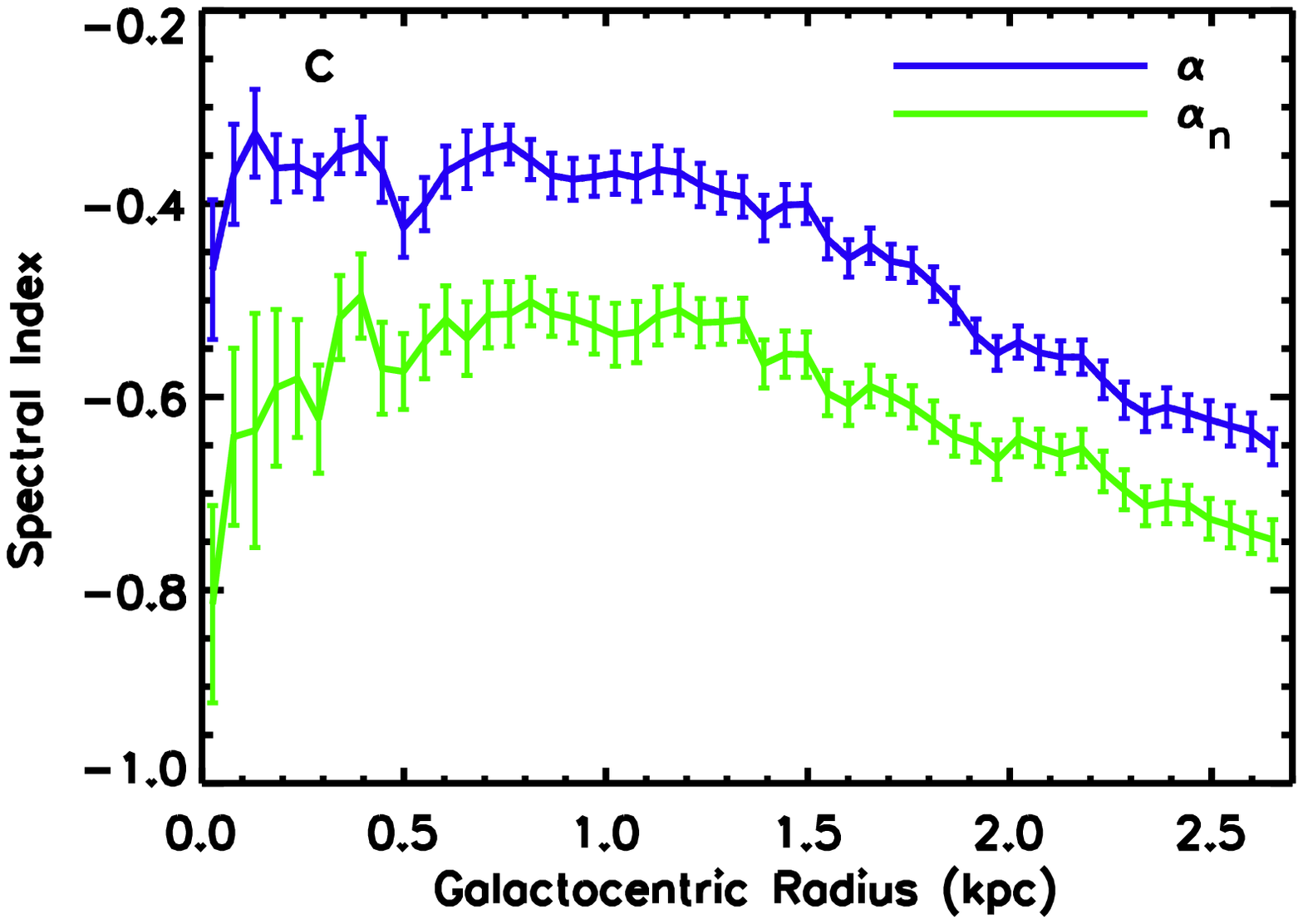}}
		\caption[]{The radial profiles of the RC emission at 1.5\,GHz  and 6.3\,GHz  and their nonthermal synchrotron and thermal free-free components ({\it a}). Also shown are the radial profiles of the thermal fraction ({\it b}) and the spectral index ({\it c}) for the observed RC emission ({\it $\alpha$}) and its nonthermal component ({\it $\alpha_{\rm n}$}). Background radio sources are subtracted. }
		\label{fig:fth}
	\end{center}
\end{figure}

\begin{figure}
	\begin{center}
		\resizebox{\hsize}{!}{\includegraphics*{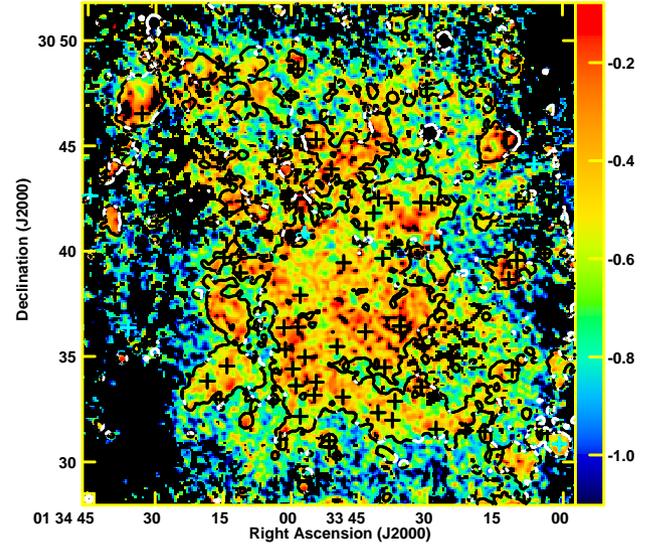}}
		\caption[]{Map of the nonthermal spectral index $\alpha_{\rm n}$ ($I^{\rm nt}_{\nu}\sim \nu^{\alpha_{\rm n}}$) obtained between 1.5\,GHz and 6.3\,GHz at 15\arcsec~ angular resolution overlaid with the 6.3\,GHz emission contours at the level of 90 $\mu$Jy/beam ($8\sigma$). Crosses show positions of the optically confirmed SNRs from \cite{Gordon_s_99}. Background radio sources are subtracted. }
		\label{fig:sp}
	\end{center}
\end{figure}

\subsection{Structure of Thermal and Nonthermal Emission}
\label{sec:tnst}

The radio source populations of M\,33 were described in Sect.~\ref{sec:pop}. The natures of the diffuse and extended components are, however, uncertain.
It is unclear what physical processes shape the observed extended features:  are they part of the gas ionized through thermal processes or are they highlighting relativistic particles travelling in the magnetized ISM? How much of this magnetized ISM is coupled with the neutral gas? Mapping the thermal and nonthermal components of the RC emission is the first step towards addressing these questions. As explained in Sect.~3.2,
the decomposition is performed at 6.3\,GHz (12\arcsec\ \footnote{Resolution is limited by the 160\,$\mu$m PACS data used to estimate the dust extinction (see Sect.~\ref{dered}).} and 15\arcsec) and 1.5\,GHz (15\arcsec resolution).
The resulting maps are shown in Fig.~\ref{fig:TNT20} at 15\arcsec resolution. 
Various structures are visible in the thermal and nonthermal maps with some similarities and differences depending on the frequency. The most striking difference occurs at 1.5\,GHz on larger scales due to the diffuse nonthermal emission, while there are some agreements on small scales. 
In particular, giant star-forming regions host both thermal and nonthermal emission, the latter in the form of SNRs and nonthermal sub-structures, although the thermal and nonthermal peak intensities can be offset from each other. At 6.3\,GHz, the diffuse nonthermal emission is less extended than at 1.5\,GHz: it is  more localized and shows a tighter correlation with the thermal emission in star-forming regions. {Figure~\ref{fig:disk}-d shows that the extended emission is more pronounced at 1.5\,GHz than at 6.3\,GHz in similar areas of the galaxy (as the colors are identical according to the spectral index, see  Table~\ref{tab:diffuse} for a more quantitative comparison).}
This can be explained by the fact that,  at higher frequencies, the synchrotron emission traces younger and more energetic CRes which are still close to their birthplace in star-forming regions. At lower frequencies, we observe the synchrotron radiation of the older CRes which are already diffused away to longer distances and to larger scales along the ISM's magnetic field.

As shown by \cite{Corbelli_20}, the 5\,GHz thermal radio sources in M\,33 represent the early phases of star formation, such as young stellar objects (YSOs) and HII regions. Many of these sources with no associated SNRs emit nonthermal emission. They also indicate the role of massive stars in the enhancement of the magnetic field and in the acceleration of cosmic ray electrons.

Table~\ref{tab:diffuse} quantifies the contribution of each structural component in the thermal and nonthermal emission. As mentioned, the 1.5\,GHz nonthermal emission emerges mainly from the diffuse disk ($\sim$~44\%) and from extended structures ($\sim$~26\%). At the higher frequency 6.3\,GHz, the contribution of diffuse emission drops, instead the nonthermal emission from discrete sources increases ($\sim$~40\%). The thermal emission is dominated by the emission from the sources at both frequencies ($\sim$~55\%). Hence, about 45\% of the thermal free-free emission must emerge from the diffuse ionized gas (DIG) with spatial scales of $>200$\,pc. This agrees with the contribution of the DIG in the total H$\alpha$ emission observed in galaxies \citep[e.g.,][]{Greenawal,Ferguson}.

In the inner $18\arcmin \times 18\arcmin$ disk of M\,33, about 40\% of the total observed flux density at 6.3\,GHz is due to thermal free-free emission, while it is 25\% at 1.5\,GHz. The origin of the observed structures differs in terms of the thermal and nonthermal radiation mechanisms as follows:
\begin{itemize}
 \item The observed diffuse disk emission, which also includes possible halo emission projected onto the disk of M\,33,  has mainly a nonthermal origin ($\sim$80\%
 at 1.5\,GHz and 66\% at 6.3\,GHz) indicating a more dominant diffusion/escape of CRes {from their birthplaces in star-forming regions} than ionizing UV photons.

  \item The observed extended structures have a similar thermal/nonthermal mixture as the diffuse disk. The extended emission follows a spiral pattern filling in the area from which no considerable thermal emission emerges (Fig.~\ref{fig:disk}) and it is dominated by the nonthermal emission.

 \item The thermal fraction of the sources is about 40\% (50\%) at 1.5\,GHz (6.3\,GHz). This is higher than that of the diffuse, extended components by a factor of two at 1.5\,GHz and is about 30\% of that found at 6.3\,GHz. The fact that more than half of the radio sources are HII regions with a flat spectrum (Sect.~\ref{sec:pop}) can explain this difference.
\end{itemize}
The synchrotron emission of the cosmic ray electrons thus shapes the diffuse and extended components of the RC emission in M\,33. This provides further hints on the distribution of the magnetic fields in the ISM traced by the synchrotron emission. The spiral pattern of the large-scale field can explain the observed extended structure of the nonthermal emission at 1.5\,GHz. This agrees with the earlier polarization studies at $\sim 0.5$\,kpc linear resolutions  \citep{Tabatabaei_2_07,Tabatabaei_08}.  The smooth nonthermal disk component can be linked to an isotropic random magnetic field component in the disk.

Compared to that at 1.5\,GHz, a newer generation of cosmic rays, which still have not diffused that far from star-forming regions, produces the synchrotron emission at 6.3\,GHz. Hence, the magnetic field structures in star-forming regions should be better traced by this emission at 6.3\,GHz.
\begin{figure}
	\begin{center}
		\resizebox{\hsize}{!}{\includegraphics*{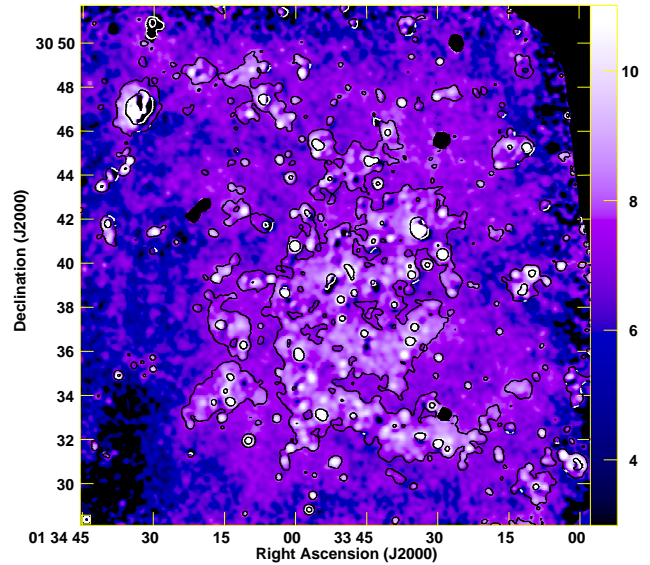}}
		\caption[]{Map of the magnetic field strength overlaid with the 6.3\,GHz emission contours at the level of 100 and 500 $\mu$Jy/beam. The angular resolution of 15\arcsec~is shown in the bottom-left corner of the image. Background radio sources are subtracted. The colour bar on the right hand side indicates the magnetic B-field strength in $\mu$G. }
		\label{fig:mag}
	\end{center}
\end{figure}

The radial profile of the RC emission obtained {by averaging the intensities} in rings of 15\arcsec~width centred at RA= 01\,h\,33\,m\,50.89s and Dec= +30\,d\,39\,m\,36.8\,s in the plane of M\,33 (i=56$^{\circ}$ and PA=23$^{\circ}$) is shown in Fig.~\ref{fig:fth}-a.  It is interesting to note that the inner most ring  shows a peak at 1.5\,GHz but not at 6.3\,GHz. The peak of the 6.3\,GHz radial profile occurs at the second ring i.e., about 80-90\,pc from the centre. Separating the thermal and nonthermal components, we infer that the 6.3\,GHz peak is actually due to the thermal emission from the young star clusters surrounding the centre and that a nonthermal source with a relatively steep spectrum (see Sect.~\ref{sec:index}) must cause the peak of the 1.5\,GHz emission at the centre. {Errors are given by the rms of the pixels in the rings divided by the square root of the number of pixels per beam area.}

The thermal fraction varies strongly from place to place in the galaxy. This is also evident from the radially averaged values in Fig.~\ref{fig:fth}-b. We note that the global decrease of the thermal fraction with $R$ is mainly due to the diffuse disk component which is about 19\% at 1.5\,GHz and 35\% at 6.3\,GHz at the centre, hence just slightly higher than the total mean value.  In star-forming regions, measurements result in thermal fractions larger than 60\% (40\%) at 6.3\,GHz (1.5\,GHz). 
\begin{figure*}
	\begin{center}
		\resizebox{\hsize}{!}{\includegraphics*{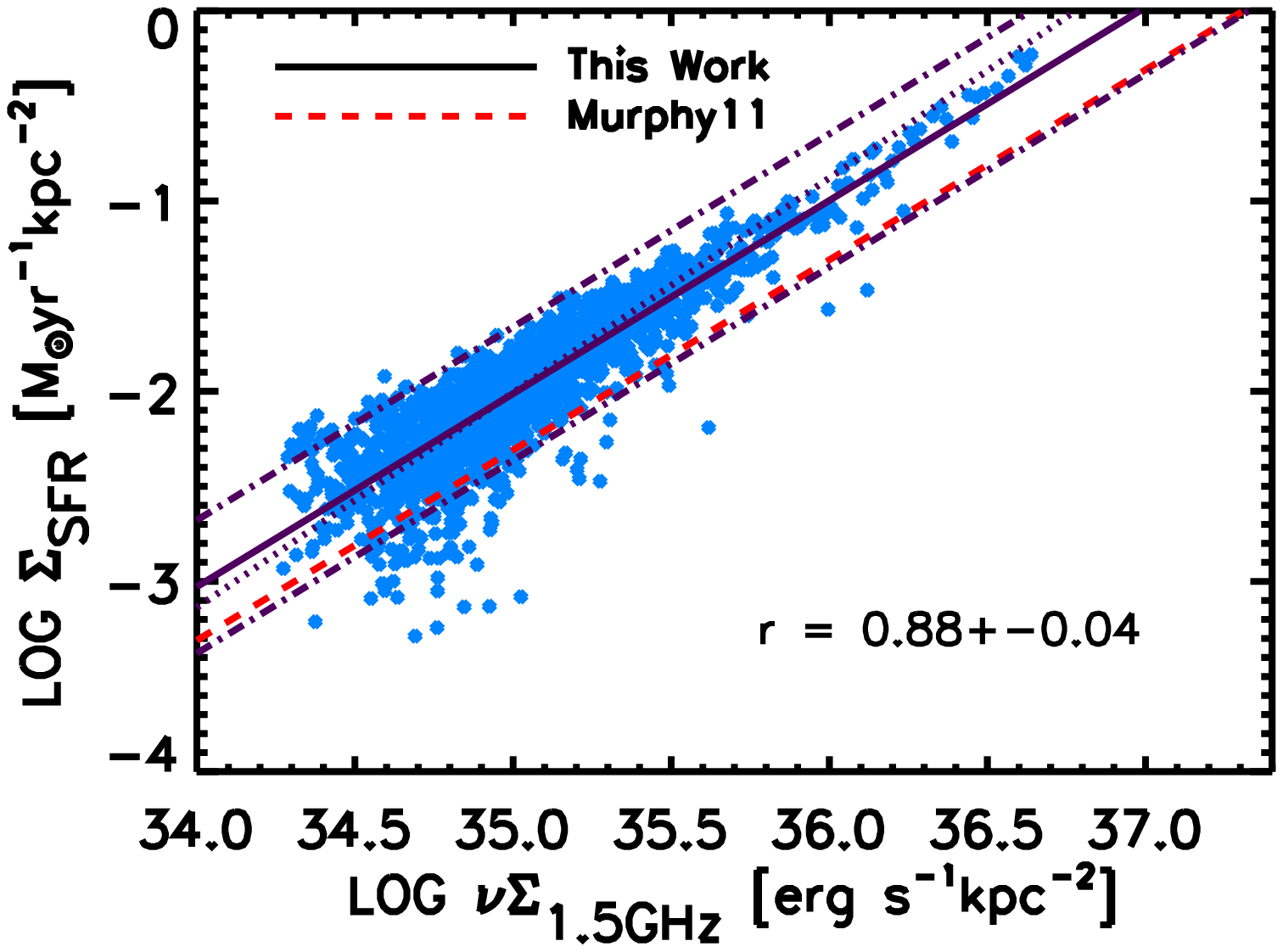}\includegraphics*{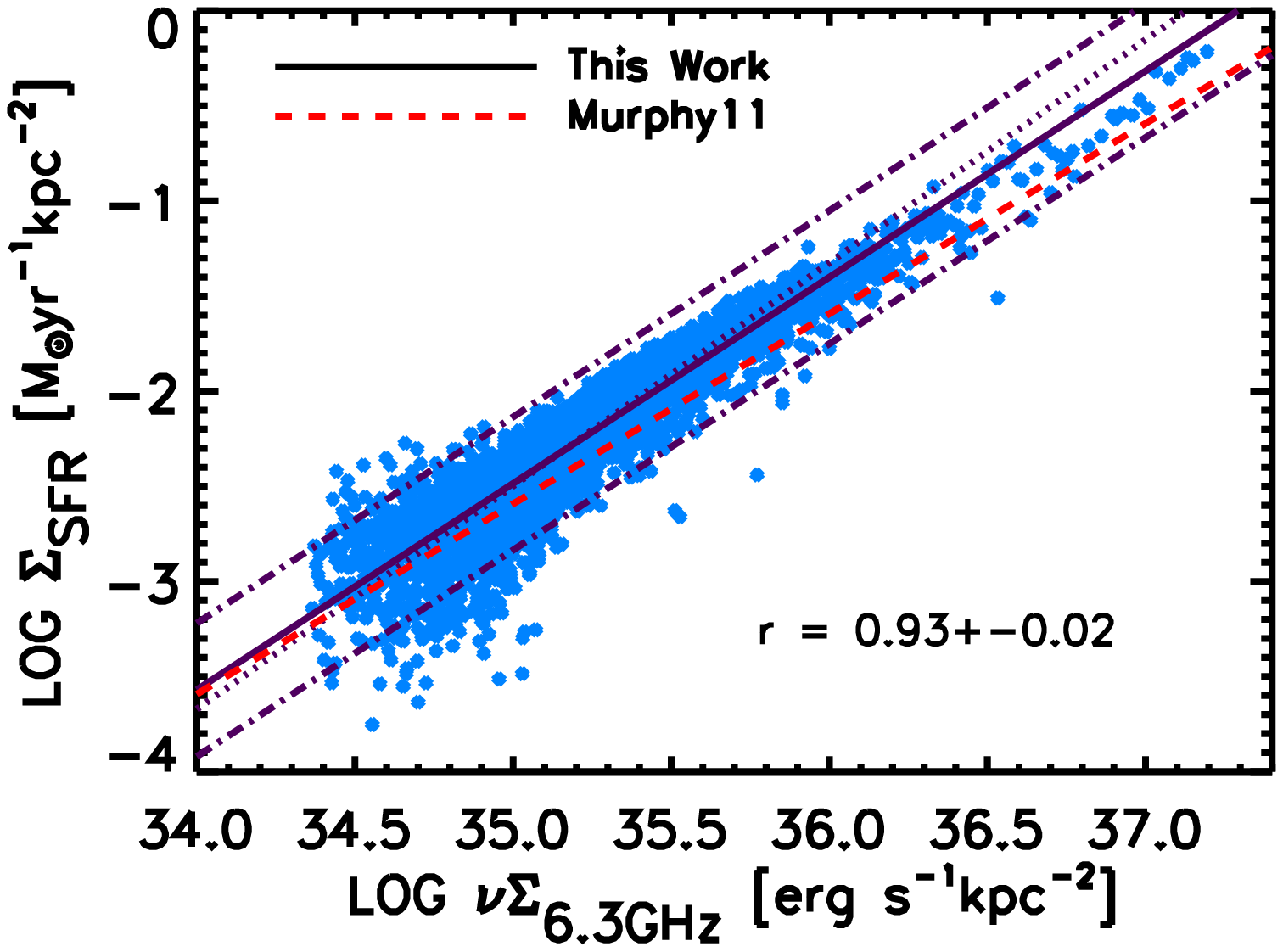}}
		\caption[]{Star-formation rate surface density $\Sigma_{\rm SFR}$ as a function of the observed RC emission at 1.5\,GHz ({\it left}) and 6.3\,GHz ({\it right}) for star-forming regions in the inner disk of M\,33. Lines show the OLS ({\it solid}) and bisector ({\it dotted}) fits and a one $\sigma$ scatter around the OLS fit ({\it dot-dashed}). Also shown is the model calibration relation ({\it red dashed line}) from \cite{Murphy_11}. The Pearson correlation coefficient r is also indicated. } 
		\label{fig:SF}
	\end{center}
\end{figure*}

\subsection{Radio Continuum Spectral Index}
\label{sec:index}
The VLA observations corrected for the short spacing and  thermal contamination are used to map the spectral index of the pure nonthermal component, $\alpha_{\rm n}$ ($I^{\rm nt}_{\nu}\sim \nu^{\alpha_{\rm n}}$). This allows one to investigate the interaction of CRes with the ISM even in diffuse regions. We measured both $\alpha$ ($I^{\rm obs}_{\nu}\sim \nu^{\alpha}$) and $\alpha_{\rm n}$ between 1.5  and 6.3\,GHz for pixels with intensities above 2.5 $\sigma$ rms noise level at each frequency. The resulting map {of $\alpha_{\rm n}$} obtained in the inner $18\arcmin \times 18\arcmin$ of M\,33 at 15\arcsec resolution is shown in  Fig.~\ref{fig:sp}.
The mean spectral index of the total RC is $\alpha=\,-0.59\pm\,0.28$, with the error the standard deviation. This is flatter than the value found for the entire galaxy M\,33 ($\alpha=\,-0.71\pm\,0.23$) but agrees with that measured in the same area by \cite{Tabatabaei_2_07}\footnote{Note that a different definition for the spectral index  and a also a different distance to M\,33 were used in that work.}.  The mean spectral index of the nonthermal emission is $\alpha_{\rm n}=\,-0.67\pm\,0.23$, slightly steeper than $\alpha$.   The actual uncertainty in the spectral index varies from 0.01 in the bright star-forming regions to 0.06 in fainter diffuse regions between the arms.

Both the total and the nonthermal spectra are flatter in star-forming complexes and generally in the spiral arms than in other parts of the disk.  The steeper spectra of the extended and diffuse structures can be linked to the CRes cooling when propagating away from their birthplace in star-forming regions due to interaction with the magnetized ISM.  Similar to the per-source measurements (Sect.~\ref{sec:pop}), the mapping approach leads to a typical value of $\alpha_{\rm n}\simeq-0.6$ for the SNRs.

Figure~\ref{fig:fth}-c shows the radial variation of the spectral index. {To generate these profiles, the mean ring spectral indices are determined from the distribution histograms of pixels in the 15\arcsec-wide rings with error bars showing the statistical errors.}  At the centre, $\alpha_{\rm n}=-0.70\pm\,0.04$. The spectrum becomes flatter moving away from the centre out to $R\sim 1$\,kpc. This is because the number of  HII complexes with flat spectra increases in this radial range. At larger radii, the spectrum steepens more or less with radius out to $R\sim 3$\,kpc, which is a diffuse region between the spiral arms, where it reaches its steepest ring average value of $\alpha_{\rm n}=-0.75\pm\,0.06$.

In the inner $18\arcmin \times 18\arcmin$ region of M\,33, $\alpha_{\rm n}$ is generally flatter than expected for pure synchrotron cooling of CRes  \citep[i.e., $\alpha_{\rm n}\simeq -1$ which prevails  in the outer parts of this galaxy,][]{Tabatabaei_3_07}. This can be explained if 1) CRes re-accelerate and gain energy due to shocks in star-forming regions and 2) they do not fully loose their energy to synchrotron loss because of advection, winds, and outflows. We further discuss these processes in Sect~\ref{sec:discuss}.

\subsection{Magnetic Field Strength}
\label{sec:mag}
The strength of the total magnetic field B$_{\rm tot}$ can be derived from the total synchrotron intensity. Assuming equipartition between the energy densities of the magnetic field and cosmic rays ($\varepsilon_{CR}\,=\,\varepsilon_{\rm B_{\rm tot}} = {\rm B}_{\rm tot}^2/8\pi$):
\begin{eqnarray}
{\rm B}_{\rm tot}= C(\alpha_n, K, L) \big[I_{\rm nt} \big]^{\frac{1}{\alpha_{\rm n}+3}},
\label{eq:Btoteq}
\end{eqnarray}
where $I_{\rm nt}$ is the nonthermal intensity and $C$ is a function of $\alpha_{\rm n}$, $K$ the ratio between the number densities of cosmic ray protons and electrons, and $L$ the pathlength through the synchrotron emitting medium  \citep[see ][]{Beck_06,Tabatabaei_08}. Using the obtained maps of $I_{\rm nt}$  and $\alpha_{\rm n}$  and assuming that the magnetic field is parallel to the plane of the galaxy (inclination of $i=56^{\circ}$ and position angle of the major axis of PA=23$^{\circ}$), B$_{\rm tot}$ is derived across the galaxy. In our calculations, we apply values of $K\,\simeq$\,100 \citep[][]{Beck_06} and  $L\,\simeq\,1\,{\rm kpc}/ {\rm cos}\,i$. 

The resulting magnetic field strength varies between 3 to 24\,$\mu$G in the inner $18\arcmin \times 18\arcmin$ region of M\,33 (Fig.~\ref{fig:mag}).  The magnetic field is stronger in star-forming complexes in the main spiral arms and in the centre than in other parts. The strongest magnetic fields are found in giant HII regions such as NGC\,604 and NGC\,595 with B$_{\rm tot} \geq 20\,\mu$G. A possible impact of massive star formation on the magnetic field strength is discussed in Sect.~\ref{sec:fb}. The mean field strength of B~$=7 \pm 1 \, \mu$G agrees with that obtained by \cite{Tabatabaei_08} in the same region, however, based on the present data we resolve much larger variations and dynamical range of the equipartion magnetic field strength.

\begin{table*}
	\begin{center}
		\caption{SFR calibrations using the RC emission in the inner $18\arcmin \times 18\arcmin$ disk of M\,33. The linear fits in logarithm scales (log\,$\Sigma_{\rm SFR}$\,~=~$b$\,\,\,log\,X + $a$) obtained using the ordinary least square (1) and bisector (2) regressions  for star-forming regions (SF) only, SF and extended structure, and the entire inner disk. Correlations refer to apertures of 15\arcsec. }
		\begin{tabular}{lllllll}
			\hline
			& X             &        \,\,\,\,\,\, $b^{(1)}$ &  $a^{(1)}$ & $b^{(2)}$  & $a^{(2)}$ & $r$ \\
			\hline
			\hline
			SF & &&& & &\\
			&$\Sigma_{\rm 1.5GHz}$ &  $1.01 \pm 0.04$  & $-37.55 \pm 0.35$ & $1.13 \pm 0.05$ & $-41.52 \pm 0.07$ & $0.88 \pm 0.04$ \\
			&$\Sigma_{\rm 6.3GHz}$ &  $1.08 \pm 0.03$  & $-40.45 \pm 0.21$ & $1.17 \pm 0.06$ &  $-43.56 \pm 0.03$& $0.93 \pm 0.02$ \\
			SF \& extended&&& && &\\
			&$\Sigma_{\rm 1.5GHz}$ &  $1.23 \pm 0.03$  & $-45.08 \pm 0.23$ & $1.54 \pm 0.04$ & $-55.99 \pm 0.04$ & $0.75 \pm 0.01$ \\
			&$\Sigma_{\rm 6.3GHz}$ &  $1.11 \pm 0.02$  & $-41.23 \pm 0.13$ & $1.22 \pm 0.02$ & $-45.10 \pm 0.02$& $0.88 \pm 0.01$ \\
			entire inner disk & &&& & &\\
			&$\Sigma_{\rm 1.5GHz}$ &  $1.60 \pm 0.02$  & $-58.48 \pm 0.23$ & $2.02 \pm 0.02$ & $-73.10 \pm 0.04$&$0.76 \pm 0.01$ \\
			&$\Sigma_{\rm 6.3GHz}$ &  $1.41 \pm 0.02$  & $-52.02 \pm 0.13$ & $1.58 \pm 0.02$ &$-57.98 \pm 0.02$ &$0.89 \pm 0.01$ \\
			\hline
			\label{tab:SFRcal}
		\end{tabular}
	\end{center}
\end{table*}

\begin{figure*}
	\begin{center}
		\resizebox{12cm}{!}{\includegraphics*{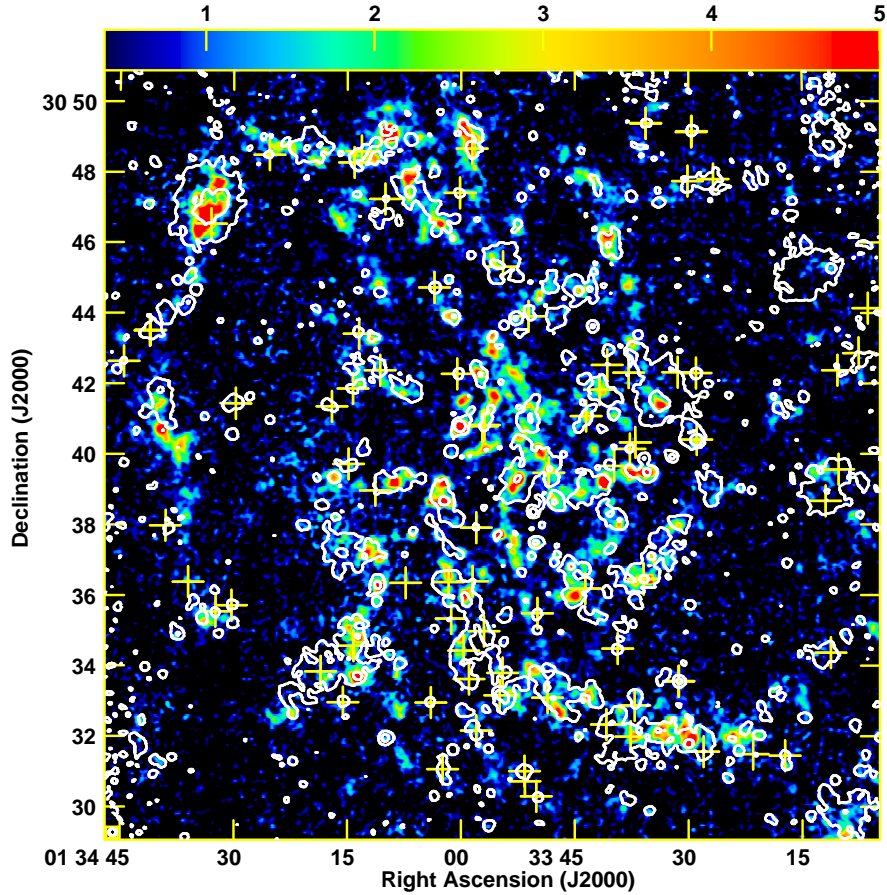}}
		\caption[]{The CO(2-1) line emission in K\,km/s with contours of the radio 6.3\,GHz emission overlaid. Contour levels are 50 and 750~$\mu$Jy per 9\arcsec~angular resolution.  Circle in the lower left corner shows the 12\arcsec~resolution of the CO(2-1) map. Crosses show positions of the optically confirmed SNRs from \cite{Gordon_s_99}. Many
			galactic SNRs are associated with giant star-forming regions.}
		\label{fig:CO}
	\end{center}
\end{figure*}

\section{Discussion}
\label{sec:discuss}
In Sect.~\ref{sec:result}, we studied the origin of the RC emission in terms of both physical processes (thermal/nonthermal) and structural components (sources/extended/disk). The structural decomposition helps in calibrating empirical recipes to derive the {star-formation rate (SFR)} surface density using the observed RC maps. In this Section, we further exploit the results of the thermal/nonthermal decomposition to study the impact of massive star formation on the two ISM components, the cosmic rays and magnetic field. Differentiating different physical processes and structures, the connection of the RC emitting ISM with its neutral gas phases is also investigated.
\subsection{Calibrating Star-Formation Rate Surface Density from Radio Continuum}
Star-forming regions are bright components of M\,33 in both  the thermal and the nonthermal maps. This is expected as these regions are the most powerful sources of the RC emission produced during different phases of massive star formation. Hence, the observed RC emission itself can be used as a dust-unobscured tracer of the SFR. However, the use of the RC maps can be complicated by the diffusion of CRes and/or the generation of secondary CRes not directly linked to star formation.  These effects become less pronounced by taking the RC emission from only star-forming regions in resolved studies. We address these effects and the impact of diffuse/extended emission on calibrating the star-formation rate surface density, $\Sigma_{\rm SFR}$, by studying the SFR--RC relation separately for star-forming regions alone (sources in Fig.~\ref{fig:disk}) and when including diffuse and extended emission in M\,33, using the results of the  structural decomposition.

The de-reddened H$\alpha$ map used as a general free-free emission template in previous sections can serve as a reference SFR tracer only after removing the emission from diffuse ionized gas (DIG) as the ionization source of the DIG can be a different one than young massive stars. However, as discussed above,  diffuse regions are also considered here to investigate their effect on the calibration. To convert the de-reddened H$\alpha$ luminosity density $\Sigma_{\rm H\alpha}$ to $\Sigma_{\rm SFR}$, we use
\begin{equation}
\left(\frac{{\Sigma_{\rm SFR}}}{M_{\sun}\,{\rm yr}^{-1} {\rm kpc}^{-2}}\right) = 5.37\times 10^{-42} \left(\frac{\Sigma_{\rm H\alpha_{\rm corr}}}{{\rm erg}\,{\rm s}^{-1}{\rm kpc}^{-2}}\right).
\end{equation}
The calibration factor is the same as that of the integrated luminosity relation given by \cite{Murphy_11}. This relation measures the current ($\lesssim$\,10 Myr) formation rate of stars assuming a Kroupa IMF \citep{Kroupa_01}.

The $\Sigma_{\rm SFR}$ and the RC luminosity surface densities  $\Sigma_{\rm 1.5\,GHz}$ and $\Sigma_{\rm 6.3\,GHz}$ are first calculated for pixels with sizes of 15\arcsec. The calibration relations are then derived taking into account a 1.5\,$\sigma$ cut off. Focusing on only star-forming regions, both the diffuse disk and the extended emission are subtracted from the radio maps\footnote{Background sources are subtracted as well.}. A tight correlation holds between $\Sigma_{\rm SFR}$ and both $\Sigma_{\rm 1.5\,GHz}$ and $\Sigma_{\rm 6.3\,GHz}$  ($r\simeq 0.90$, see Fig.~\ref{fig:SF}) in the star-forming regions. Fitting a relation of the form log\,$\Sigma_{\rm SFR}$\,~=~$b$\,\,\,log\,X + $a$, 
a linear relation is found between $\Sigma_{\rm SFR}$ and $\Sigma_{\rm 1.5\,GHz}$ using an ordinary least square (OLS) regression (see Table~\ref{tab:SFRcal}).
The correlations become steeper, particularly at 1.5\,GHz, if the extended structures are also included in this process. If neither the extended emission nor the diffuse disk is subtracted, the slopes of the $\Sigma_{\rm SFR}$ relations increase by  $\simeq$~60\% and 30\% vs. $\Sigma_{\rm 1.5\,GHz}$ and $\Sigma_{\rm 6.3\,GHz}$, respectively\footnote{This steepening is even higher using the bisector regression.}. In this case, the super-linearity of the $\Sigma_{\rm SFR}$--$\Sigma_{\rm 1.5\,GHz}$ correlation is similar to that obtained by \cite{Heesen_14} for  the SINGS galaxies (their reported average slope of 0.63 of the RC vs. $\Sigma_{\rm SFR}$ is equivalent to a slope of $\simeq 1.6$ for the $\Sigma_{\rm SFR}$ vs. $\Sigma_{\rm 1.5\,GHz}$ relation). This shows the impact of diffuse emission in resolved studies of the SFR--RC correlation. The inclusion of a galaxy's diffuse regions poses a larger weight to the low-SFR regions, where the RC luminosity is in enhanced due to diffused CRes or generated in secondary processes. Because these CRes have lower energies, this effect is more pronounced at lower radio frequencies ($\nu_{\rm c}\sim E^2$). This explains the different SFR--RC slopes at 1.5 and 6.3\,GHz if the extended/diffuse emission is not subtracted.

Focusing only on the star-forming regions, we find the following calibration relation at 1.5\,GHz,
\begin{eqnarray}\label{eq:cal1}
\left(\frac{{\Sigma_{\rm SFR_{1.5}}}}{M_{\sun}\,{\rm yr}^{-1} {\rm kpc}^{-2}}\right)\, & = & \,\left. (2.81 \pm 0.98) \,\times 10^{-38} \right. \\
& & \left. \times \left(\frac{\nu \Sigma_{\rm 1.5\,GHz}}{{\rm erg}\,{\rm s}^{-1}{\rm kpc}^{-2}}\right)^{(1.015\pm 0.042)}, \right. \nonumber
\end{eqnarray}
and at 6.3\,GHz,

\begin{eqnarray}\label{eq:cal2}
\left(\frac{{\Sigma_{\rm SFR_{6.3}}}}{M_{\sun}\,{\rm yr}^{-1} {\rm kpc}^{-2}}\right) \, & = & \,\left. (3.55\pm 0.76)\times 10^{-41} \right. \\
& & \left. \times \left(\frac{\nu \Sigma_{\rm 6.3\,GHz}}{{\rm erg}\,{\rm s}^{-1}{\rm kpc}^{-2}}\right)^{(1.085\pm 0.035)}, \right. \nonumber
\end{eqnarray}
The $\Sigma_{\rm SFR}$--$\Sigma_{\rm 6.3\,GHz}$ relation is slightly steeper than the $\Sigma_{\rm SFR}$--$\Sigma_{\rm 1.5\,GHz}$ one, though they overlap considering the errors. 
The empirical calibration relations given in Eqs.~\ref{eq:cal1} and \ref{eq:cal2} are shown as solid lines in Fig.~\ref{fig:SF}. Assuming a linear proportionality between supernova rate and the SFR and using an empirical  Galactic relation between the nonthermal spectral luminosity and the supernova rate, \cite{Murphy_11} (hereafter Murphy11) presented a model calibration that depends on the nonthermal spectral index (their Eq.~15). We also obtain  $\Sigma_{\rm SFR_{\nu}}$ using Murphy's calibration (dashed lines in Fig.~\ref{fig:SF}) by adopting M\,33's average $\alpha_{\rm n}=-0.67$ in the inner disk (see Sect.~4.3). This calibration, however, underestimates M\,33's SFR by a factor of $\simeq$\,2 and 1.5 derived from our 1.5\,GHz and 6.3\,GHz data, respectively.
As explained by Murphy11, such underestimates are linked to the diffusion or escape of CRes neglected in the model. We note, however, that the deviation is within one $\sigma$ scatter of the fitted empirical relation.

\begin{figure}
	\begin{center}
		\resizebox{\hsize}{!}{\includegraphics*{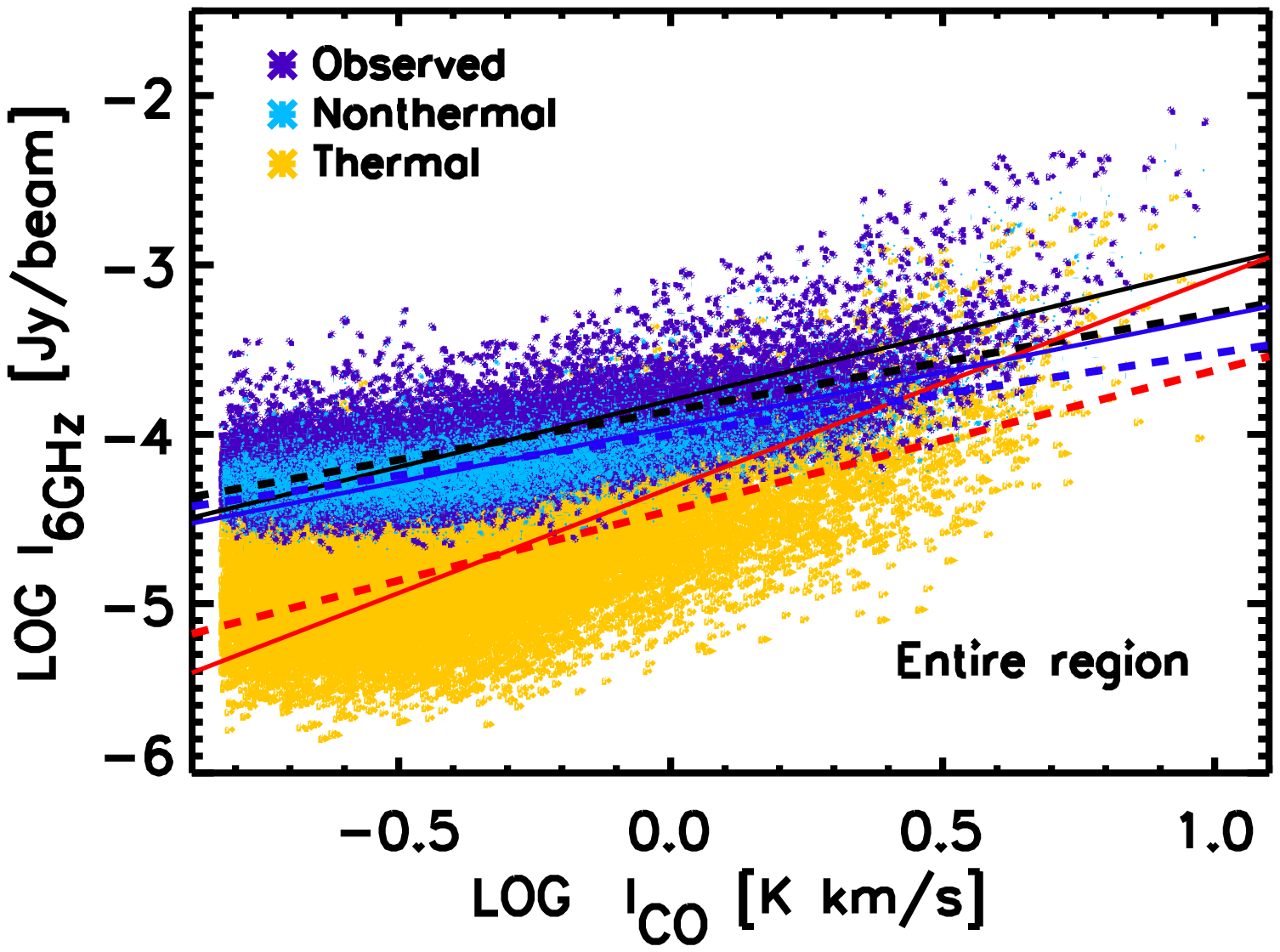}}
		\resizebox{8cm}{!}{\includegraphics*{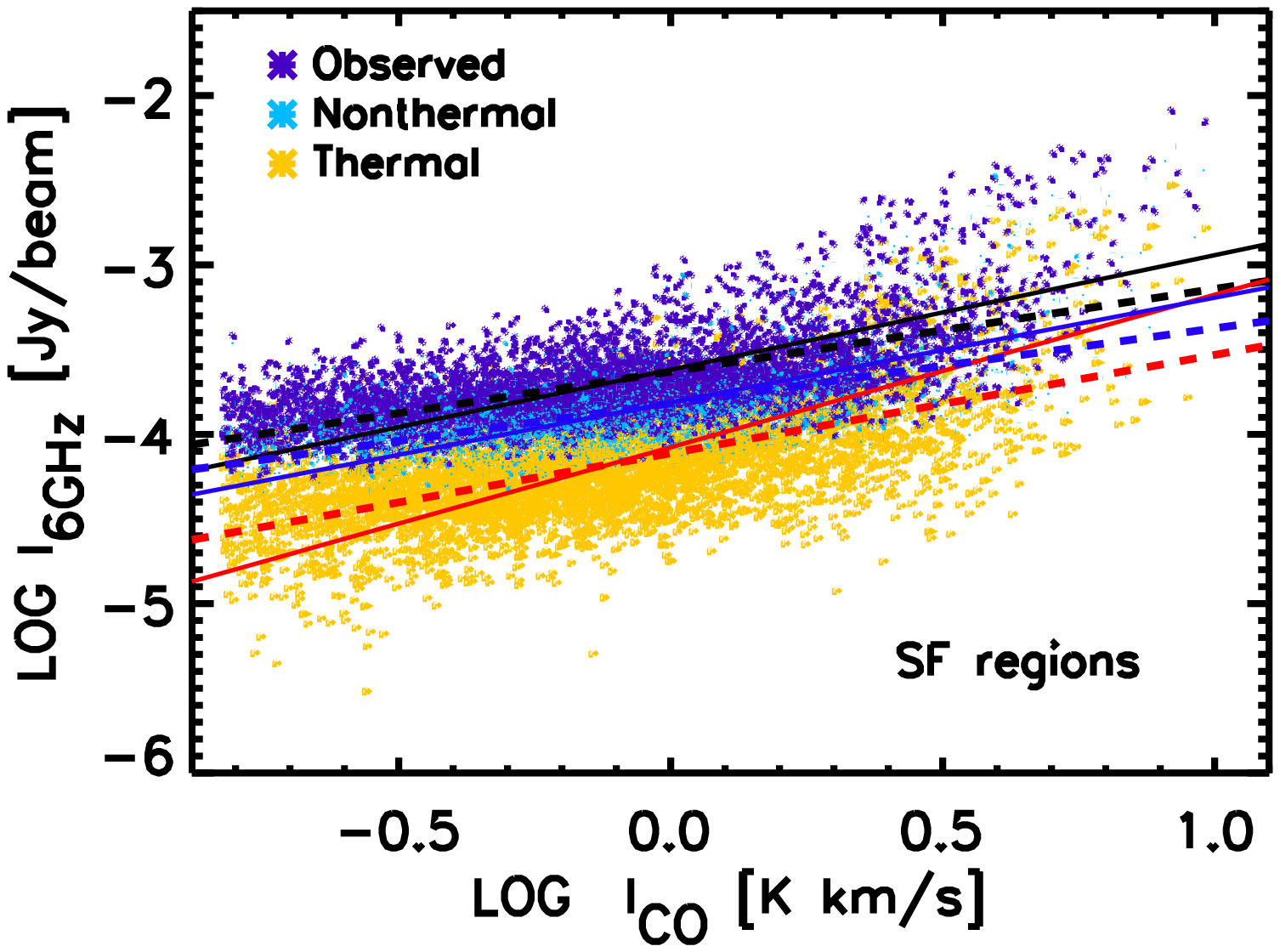}}
		\resizebox{8cm}{!}{\includegraphics*{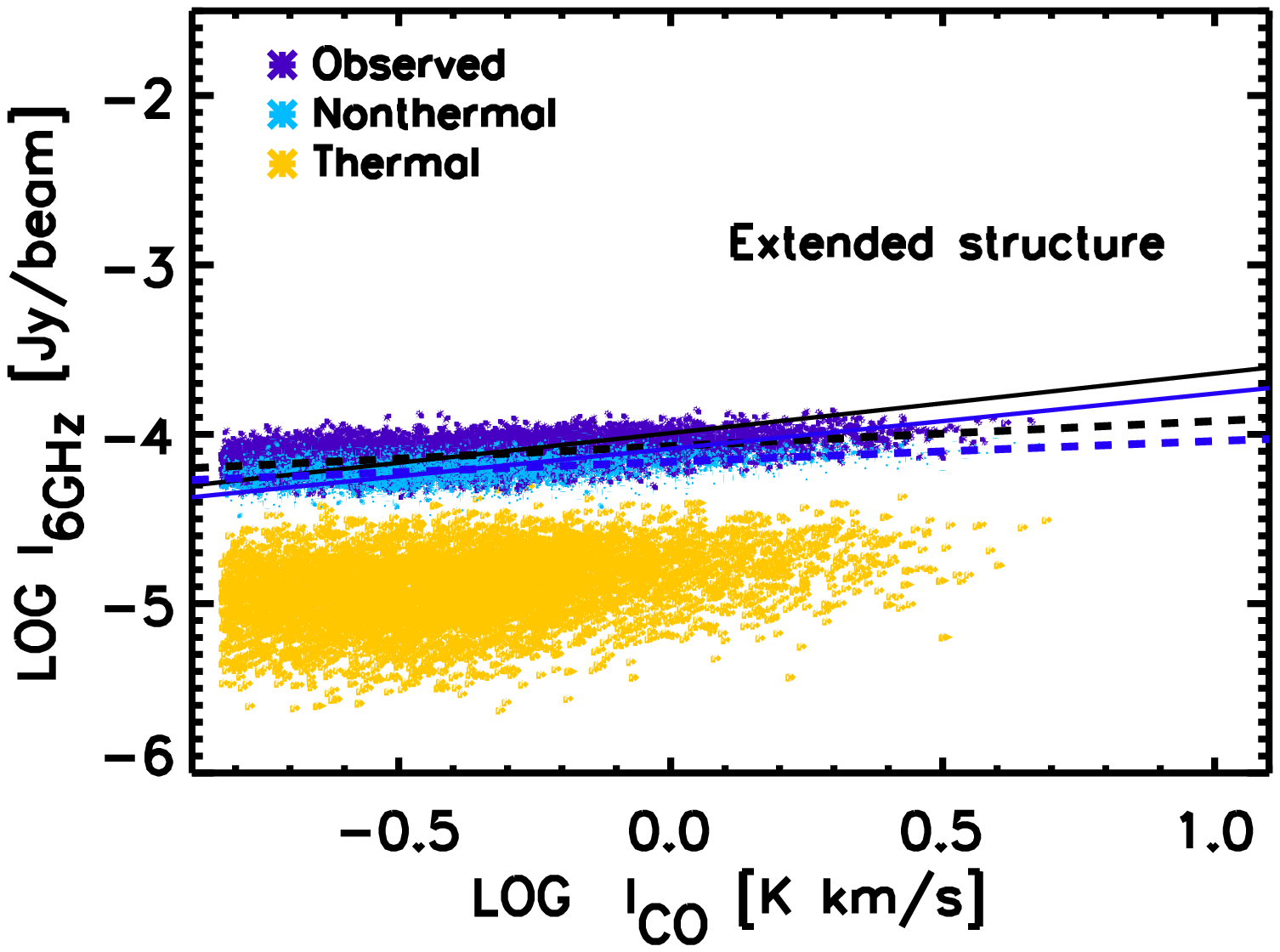}}
		\caption[]{The 6.3\,GHz RC correlation with the molecular gas traced by CO(2-1) line emission for the observed, nonthermal, and thermal RC emission of the entire inner 18\arcmin $\times$18\arcmin\ disk of M\,33. Lines show the OLS bisector ({\it solid}) and OLS ({\it dashed}) fits for the observed ({\it black}), nonthermal ({\it blue}), and thermal ({\it red})  RC--CO correlations. }
		\label{fig:CORC}
	\end{center}
\end{figure}

\subsection{Comparison with Molecular Gas}
 Studying the correlation between the RC emitting processes and the neutral gas is the first step towards understanding the pressure balance in the multi-phase ISM. The thermal RC component, which traces the ionized gas and can be used as a reference SFR tracer (e.g., Murphy11), is expected to correlate with neutral gas emission according to the Kennicutt-Schmidt (KS) star-formation law \citep{kennicutt_98}. The nonthermal RC component, which traces the relativistic and magnetized ISM, is also stronger in star-forming regions and hence is expected to correlate with the neutral gas as well. However, it is unclear whether the correlations for the non-thermal component are limited to star-forming regions only, particularly  as there is observational evidence showing a tight RC--molecular gas correlation which cannot be linked to massive star formation \citep{Schinnerer_13,Taba_18}. This motivates us to investigate these correlations in M\,33, separately in star-forming regions and in other parts of the ISM.

 Figure~\ref{fig:CO} shows that the RC emission agrees generally with the molecular gas traced by CO(2-1) line emission in most star-forming complexes. Apart from this agreement, it is interesting to note that the correlation drops at the locations of SNRs: The CO emission drops to below a 3 sigma level in most cases, particularly, for those SNRs indicated by crosses. 
  Hence, a feedback from the SNRs is likely responsible for clearing the molecular gas in their surroundings.

 The CO emission is not only weak in the presence of the SNRs. For instance, no CO emission is found in IC131, another giant HII region, hosting no SNR. This source is however known to be a strong X-ray emitter \citep{Tullman}.

 Quantifying the correlation in the inner disk of M\,33, the RC and CO maps are cross-correlated after removing the radio BG sources and convolving to the same angular resolution.   Table~\ref{tab:rcgas} lists the correlation coefficients and the RC vs. CO power-law slopes obtained through bisector fits\footnote{Since the correlated variables do not directly depend on each other, we fitted a power law to the bisector in each case \citep{Isobe,Hoernes_etal_98}. }. Errors in the Pearson correlation coefficients are given by $\Delta r_{c}= \sqrt{1-r^{2}_{c}}/ \sqrt{n-2}$, with $n$ being the number of pixels in an image.
 In the star-forming regions, we find a modest RC--CO correlation with a Pearson coefficient of $r^{\rm SF}\simeq0.60$ that is similar at both frequencies (slightly tighter with the nonthermal RC, Fig.~\ref{fig:CORC}) at the pixel resolution of 15\arcsec. %
 The observed RC  emission changes with the CO emission through a sub-linear relation with an exponent of $b^{\rm SF}=0.67\pm 0.02$. A similar relation holds between the nonthermal RC and CO. However, the thermal--CO correlation is almost linear and agrees with the molecular K-S relation in the SINGS sample of galaxies \citep{Bigiel_08}.
 This correlation deviates from linearity ($b\simeq1.25$) when including also other regions of the ISM in the inner 18\arcmin\ $\times$18\arcmin\ disk, while the quality of the correlation remains unchanged. A super-linear correlation in M\,33 was already reported by \cite{Heyer}, \cite{Verley_10}, \cite{Williams_18} showing the mixing effect of diffuse and SF-related emission and the importance of performing structural decomposition in resolved studies.

 In regions where the RC emission has an extended/diffuse structure (indicated as 'extended' in Table~\ref{tab:rcgas}), no significant correlation holds between the thermal RC and the molecular gas $r^{\rm extended}\simeq0.30$, while the observed and nonthermal RC emission are still correlated with it, $r^{\rm extended}\simeq0.45$. { Hence, the RC--CO correlation is not solely due to massive star formation as it can also hold in more quiescent regions of the ISM {that} is less efficient in forming those massive stars. A fine balance between the magnetic fields, CRes, and molecular gas must then hold in these regions \citep[see e.g.,][]{Niklas_977,Murgia,Taba_13}, irrespective of the origin of the magnetic field. More detailed studies are needed to dissect the origin of the magnetic fields which are important in the observed nonthermal RC--CO correlation. While in star-forming regions, a correlation between turbulence and magnetic fields amplified by a small-scale dynamo can play an important role in preserving the RC--CO correlation, this process is likely not as efficient in more quiescent ISM (see Sect.~\ref{sec:fb}). Investigating the structure of the magnetic field traced in polarization helps understanding its amplification mechanisms. This will be further addressed in a forthcoming paper.

 \cite{Berkh} showed that the correlation between the nonthermal RC  and dust emission flattens due to diffusion of CRes. {This} can explain the relatively flat nonthermal RC--CO correlation obtained in the extended/diffuse RC emitting regions ($b^{\rm extended}=0.29$ compared to $b^{\rm SF}=0.67$, Table~\ref{tab:rcgas}) as there the population of CRes is dominated by those {that} are diffused away from star-forming regions. }
\begin{figure}
	\begin{center}
		\resizebox{\hsize}{!}{\includegraphics*{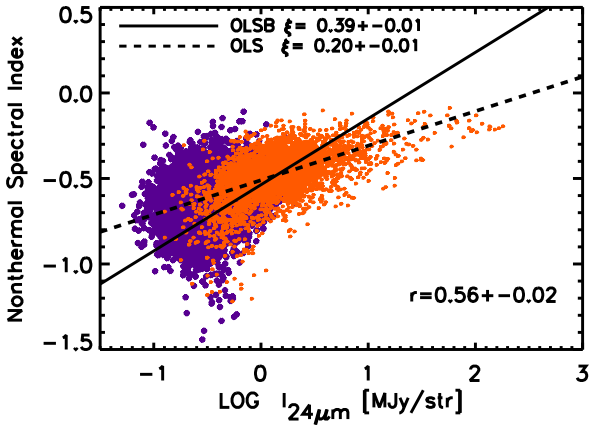}}%
		\caption[]{ The nonthermal spectral index (obtained above a 2.5\,$\sigma$ cut-off) becomes flatter with $\Sigma_{\rm SFR}$ traced by the 24$\mu$m emission in star-forming regions ({\it orange}). Lines show the OLS bisector (OLSB, solid) and OLS (dashed) regressions. The Pearson correlation coefficient $r$ is also indicated. No significant correlation holds in the extended/diffuse regions ({\it violet points}). }
		\label{fig:alpha_sf}
	\end{center}
\end{figure}

The resolved RC--CO correlation in more massive spiral galaxies, such as M51, is found to be tighter than in M\,33 \citep{Paladino,Schinnerer_13}. This can be linked to a different molecular gas fraction and/or different physical conditions of the gas in these galaxies \citep[e.g., mid-plane pressure; ][]{Mac}. The ISM is dominated by the atomic HI gas in M\,33, unlike in M51 \citep{Scoville_83}. As shown by \cite{Druard}, the molecular  H$_2$ forms quiescently from the denser atomic clouds in M\,33. Investigating the correlation between the RC and the HI-21\,cm line emission in M\,33, we find a large scatter ($r<0.33$) and the correlations with the total neutral gas (HI+H$_2$) are weaker than those with the molecular gas. This is expected if most of the RC emission emerges from the star-forming regions in the molecular phase rather than in the atomic phase. 
\begin{table*}
	\begin{center}
		\caption{Correlation between the molecular gas and observed (OBS), nonthermal (NTH), and thermal (TH) RC emission in star-forming regions ($r^{\rm SF}$), remaining parts of the ISM ($r^{\rm extended}$), and the entire ISM ($r$) in the inner 18\arcmin $\times$18\arcmin\ region of M\,33. The corresponding power-law exponent $b$ in the RC\,$\propto {\rm I_{CO}}^b$ relation fitted is obtained using the bisector regression. Correlations  refer to apertures of 15\arcsec.}
		\begin{tabular}{ l l l l l l l}
			\hline
			RC     &  $r^{\rm SF}$ &  $b^{\rm SF}$ & $r^{\rm extended}$ & $b^{\rm extended}$ &$r$ & $b$ \\
			\hline
			\hline
			1.5GHz &&&&& &\\
			OBS & 0.63\,$\pm$\,0.02   & 0.67\,$\pm$\,0.02 & 0.45\,$\pm$\,0.01 & 0.29\,$\pm$\,0.02 & 0.61\,$\pm$\,0.01 & 0.41\,$\pm$\,0.02\\
			NTH& 0.63\,$\pm$\,0.02   &  0.65\,$\pm$\,0.04 & 0.40\,$\pm$\,0.01 &0.29\,$\pm$\,0.02 & 0.60\,$\pm$\,0.01 & 0.50\,$\pm$\,0.02\\
			TH& 0.59\,$\pm$\,0.02  & 0.96\,$\pm$\,0.04  & 0.31\,$\pm$\,0.01  & ...........& 0.59\,$\pm$\,0.01& 1.25\,$\pm$\,0.04\\
			\hline
			6.3\,GHz & & & & & &\\
			OBS & 0.62\,$\pm$\,0.01  & 0.68\,$\pm$\,0.02  & 0.49\,$\pm$\,0.01 & 0.35\,$\pm$\,0.02&0.62\,$\pm$\,0.01 &0.79\,$\pm$\,0.02 \\
			NTH& 0.62\,$\pm$\,0.01  & 0.62\,$\pm$\,0.02  & 0.46\,$\pm$\,0.01 &0.33\,$\pm$\,0.02 & 0.62\,$\pm$\,0.01 &0.65\,$\pm$\,0.02\\
			TH&  0.57\,$\pm$\,0.01  &  0.90\,$\pm$\,0.05 & 0.33\,$\pm$\,0.01 &........... & 0.59\,$\pm$\,0.01 & 1.24\,$\pm$\,0.04\\
			\hline
				
		\end{tabular}
		\label{tab:rcgas}
	\end{center}
\end{table*}

\begin{figure*}
	\begin{center}
		\resizebox{\hsize}{!}{\includegraphics*{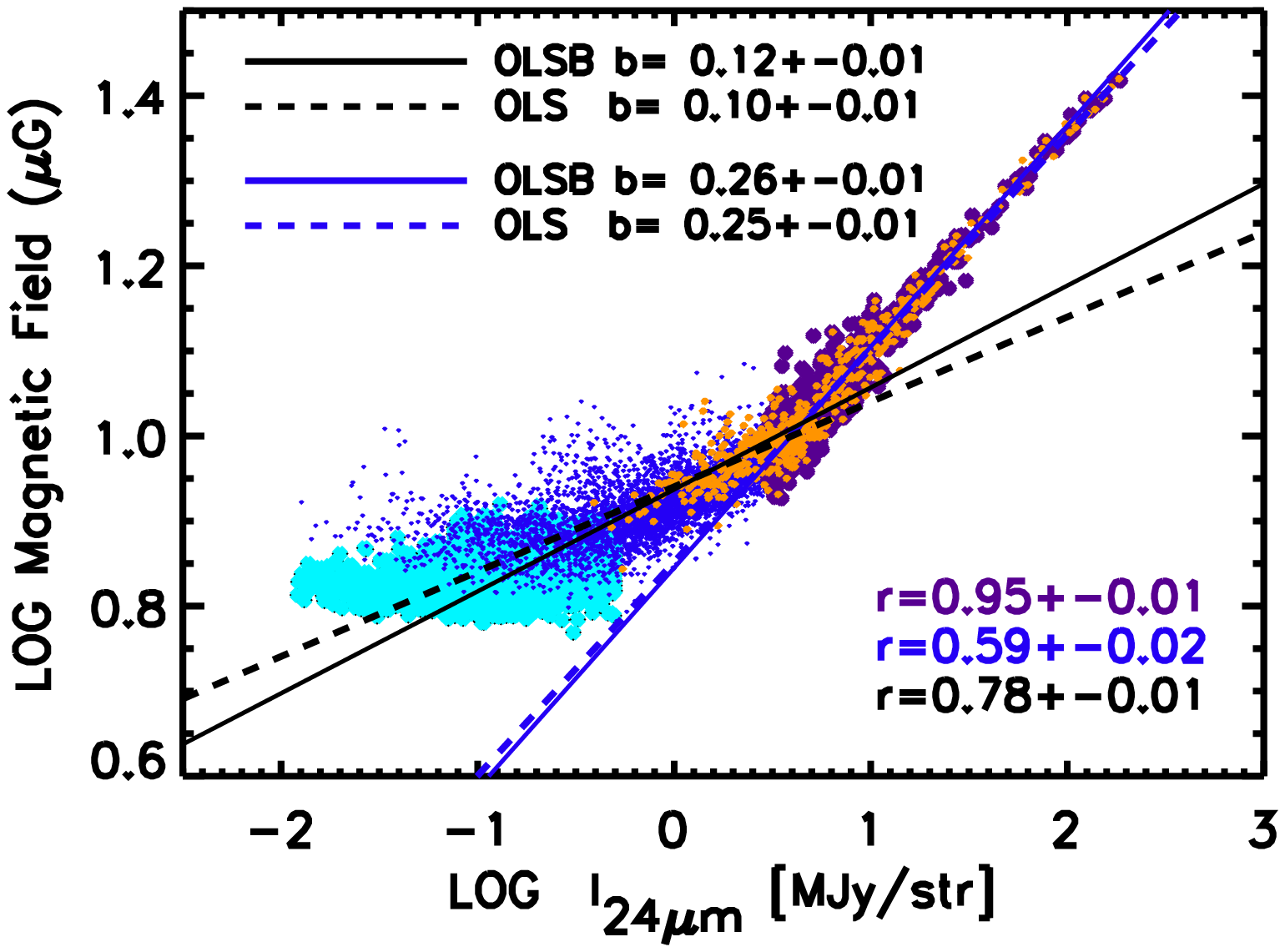}\includegraphics*{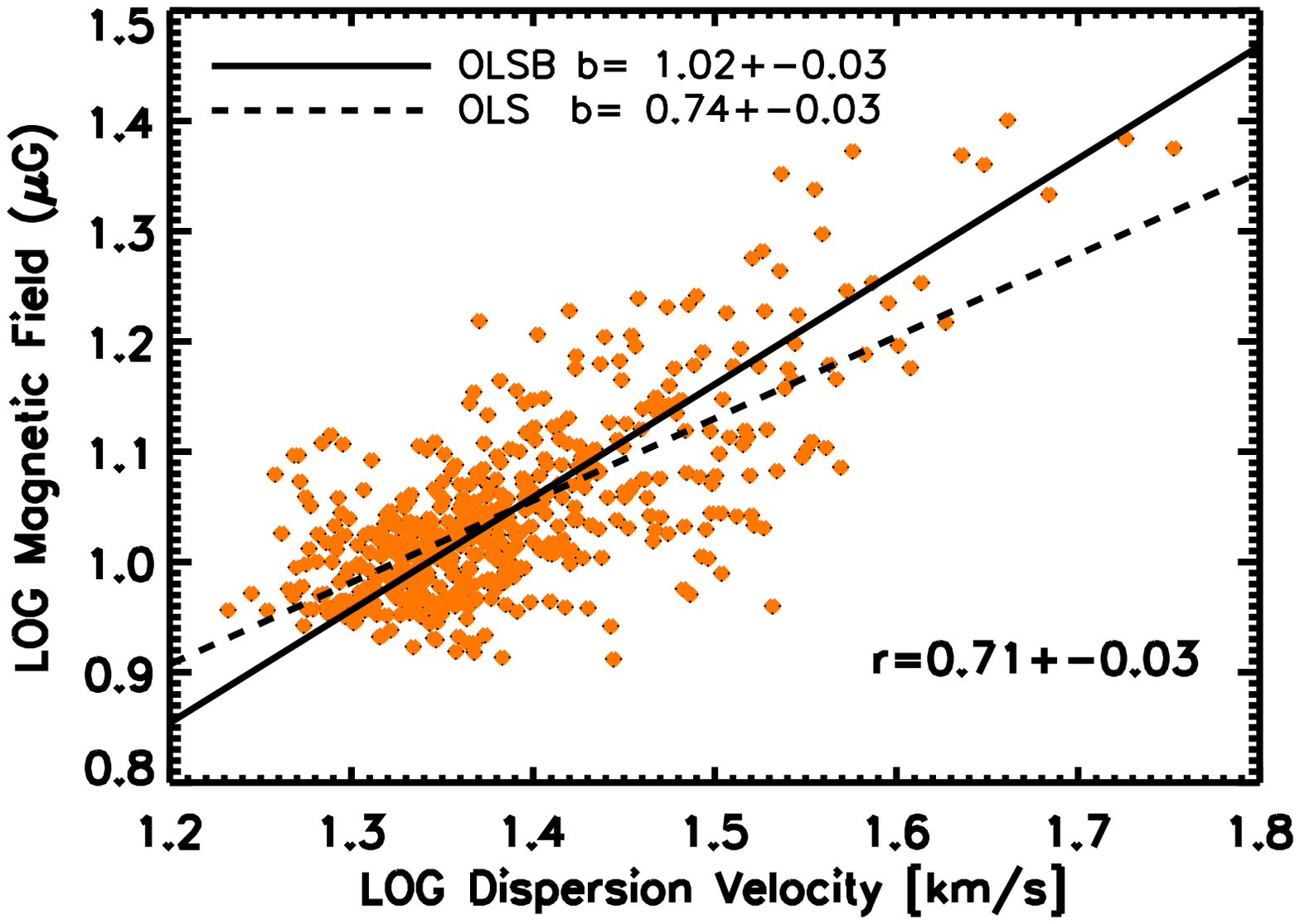}}%
		\caption[]{{\it Left:} the magnetic field strength vs. $\Sigma_{\rm SFR}$ traced by the 24$\mu$m emission. A bi-modal behavior is found at I$_{\rm 24\mu m}\geq \,3\, {\rm MJy sr^{-1}}$ ({\it magenta points}) and below ({\it blue points}), i.e., depending on the SFR regime. Lines show the fits to the entire data ({\it black}) and the high-SFR regime ({\it blue}). No significant correlation is found in the extended/diffuse region ({\it cyan points}).  {\it Right:} the magnetic field strength increases with the gas turbulent velocity (of H$\alpha$ lines) in a sample of bright star-forming regions. These regions are shown in the left panel ({\it orange points}) as well. Lines show the OLS bisector (OLSB) and OLS regressions. The Pearson correlation coefficients $r$ are also indicated.   }
		\label{fig:b-sfr}
	\end{center}
\end{figure*}

\subsection{Impact of Star Formation on CRes and Magnetic Field}
\label{sec:fb}
Massive star-formation activity can change the physical state of the surrounding ISM by heating, ionization, and powerful winds. It can also inject turbulence, as well as  high-energy particles and CRes with a flat spectrum. The latter can be investigated by mapping the pure synchrotron spectral index $\alpha_{\rm n}$ in galaxies. This map obtained for M\,33 (Sect.~4.3) already indicates that the CRe energy spectrum is flatter in M\,33's giant HII regions (Fig.~\ref{fig:sp}).  We further investigate if there is an overall trend between $\alpha_{\rm n}$ and $\Sigma_{\rm SFR}$ in this galaxy. For this purpose, we use a different SFR tracer, the 24\,$\mu$m emission\footnote{The use of  RC or H$\alpha$ emission as SFR tracer may inject artificial correlations for this specific comparison.}. A correlation is found between $\alpha_{\rm n}$ and the 24\,$\mu$m intensity, I$_{\rm 24\mu m}$, with a Pearson coefficient of $r=0.57\,\pm\,0.02$  (Fig.~\ref{fig:alpha_sf}). For star-forming regions, i.e., after subtracting the diffuse emission,  we find the following  relation for 15\arcsec~pixels\footnote{$\alpha_{\rm n}$ is re-calculated using intensity maps with 15\arcsec\~pixels.}
\begin{equation}
\alpha_{\rm n}= \eta + \zeta\,.\, {\rm log \left(\frac{ I_{\rm 24\mu m}}{MJy\,sr^{-1}}\right)},
\end{equation}
with $\eta=-0.536\pm 0.003$ and $\zeta=0.39\pm0.01$ using the bisector regression. We note that a flatter variation is found, $\zeta=0.21\pm0.01$, using the OLS fitting. This shows that the flattening of the synchrotron spectrum with star formation is a general trend in the inner disk of M\,33. The slope of variation $\zeta$ agrees with our global finding in the KINGFISHER \citep[Key Insight in Nearby Galaxies Emitting in Radio,][]{Taba_17} sample of galaxies: galaxies with higher $\Sigma_{\rm SFR}$ have a flatter synchrotron spectrum. This flattening was linked to the pitch-angle scattering of high-energy CRes in turbulent magnetic fields that is likely stronger in galaxies with higher SFRs. This might also explain the observed trend in M\,33, as the magnetic field strength B increases with $\Sigma_{\rm SFR}$. Figure~\ref{fig:b-sfr} (left panel) shows B vs. $\Sigma_{\rm SFR}$ as traced by the 24\,$\mu$m intensity, plotted using a 5$\sigma$ rms cut. A best fit OLS bisector regression in the log-log plane results in the following relation
\begin{equation}
{\rm log} \left(\frac{\rm B}{\rm \mu G}\right)= a\, +\, b\times {\rm log \left(\frac{ I_{\rm 24\mu m}}{MJy\,sr^{-1}}\right)},
\end{equation}
with $a=0.94\pm 0.01$ and $b=0.12\pm 0.01$ and a Pearson correlation of $r=0.76\pm 0.01$ (black lines in Fig.~\ref{fig:b-sfr} (left). However, a steeper variation is inferred at star-formation rates higher than a certain level. In other words, the power-law proportionality of  B~$\propto \Sigma_{\rm SFR}^b$ has a bi-model variation with
\begin{eqnarray}
b&=& \left. 0.26 \pm 0.01,  \,\,\,\,\,\,\,\,\,\,\,  {\rm I_{24\mu m}}\geq 3\,{\rm MJy\,sr^{-1}}, \right. \nonumber \\
b& =& \left. 0.10 \pm 0.01,     \, \,\,\,\,\,\,\,\,\,\,  {\rm I_{24\mu m}}< 3\,{\rm MJy\,sr^{-1}}, \right. \nonumber
\end{eqnarray}
where the fits use the OLS bisector regression (indicated as OLSB in Fig.~\ref{fig:b-sfr}).
The power-law index of $\simeq$0.26 obtained is close to the theoretical value of 0.3  proposed for amplification of the magnetic field by a small-scale dynamo in SF regions \citep[][]{Schleicher}. This mechanism seems, however, not to be efficient in amplifying B at low star-formation rates as indicated by the flatter power-law index of $\simeq$0.10. %
If the magnetic field in SF regions is mainly amplified by the action of turbulence, a positive correlation is expected between B and gas turbulent velocity. We further investigate this by using the H$\alpha$ velocity dispersion measured for bright star-forming regions of M\,33. The right panel in Fig.~\ref{fig:b-sfr} indeed shows a positive and linear correlation (OLSB fit) between the magnetic field strength and the turbulent velocity. Hence, star-formation feedback increases the turbulent velocity of its surrounding gas. This turbulent energy is converted into magnetic energy, e.g., due to the dynamo action. This results in a turbulent and tangled magnetic field enhancement in the ISM.

As already mentioned in Sect.~1, previous studies show a relatively flat nonthermal spectrum in complexes of star-forming regions where the magnetic field is strong \citep{Tabatabaei_3_07,Taba_13,Hassani}, in agreement with our findings. This may seem counter-intuitive because synchrotron loss should increase with increasing magnetic field strength, assuming that the field is uniform. However, our studies show that the magnetic field is strongly tangled and non-uniform in star-forming complexes and, hence, CRes can experience pitch angle scattering \citep[e.g.,][]{tautz}. These energetic CRes tend to stream in related dense plasma at a speed larger than the local Alfv\'{e}n speed {(which drops with density $\rho$ as $v_{\rm A}=B/ \sqrt{4 \pi \rho}$)} causing streaming instabilities \citep[e.g.,][]{Marco}. As a result, the synchrotron emission observed in star-forming complexes is due to CRes {with a short residence time in regions of strong turbulent magnetic fields}. This results in a relatively flat spectrum because CRes do not have enough time to cool down efficiently by the magnetic field. These energetic CRes can have an important consequence for the energy balance of the ISM in star-forming regions causing strong winds and outflows as a result of a pressure gradient imposed by them. The existence of a nonthermal halo in M\,33, indicated through a detailed study of the radio-IR correlation \citep{Taba_13_b}, is consistent with such a feedback. Observations of edge-on galaxies also agree with the cosmic-ray-driven winds related to star-forming regions in galactic disks \citep{Heesen_18}.

We note that a flat synchrotron spectrum can also be maintained through other processes in the star-forming regions. As shown in Fig.~\ref{fig:sp}, a fraction of the flat-spectrum regions appears as shells around the SNRs and giant HII complexes (with $\alpha_{\rm n}$ $>-0.5$). Such features can be produced if CRes have gained additional energy due to re-acceleration in related shock fronts. Taking into account the resolution (pixel scale of $\leq$20pc) of this study, detecting these features is not far from expectation, particularly in large complexes of star-forming regions in M\,33.

\section{Summary}

We study the structure and origin of the RC emission from the ISM of M\,33 down to linear scales of $\simeq$~30\,pc (9\arcsec.3) and 50\,pc (15\arcsec) by combining  VLA interferometric observations at C and L bands with  single-dish observations from the 100-m Effelsberg telescope. The RC emission emerges from three main structural components, bright star-forming regions, extended structures, and a diffuse disk. Separating these components helps in calibrating the star-formation rate and addressing the effect of diffuse emission when using the RC maps.  We also disentangle  the thermal and nonthermal processes in the star-forming and diffuse parts of the gaseous ISM.  This helps us to investigate the impact of massive star formation on the magnetic field and CRes and further resolve the nature and origin of feedback in M\,33. The most important results and conclusions are summarized as follows.
\begin{itemize}
\item Radio sources contribute about 36\% (46\%) of the total RC emission at 1.5\.GHz (6.3\,GHz) in the inner 18\arcmin $\times$18\arcmin\ disk of M\,33 at 15\arcsec\ (50\,pc) resolution. About half of these sources are HII regions (contributing 57\% and 74\% in total source emission at 1.5\,GHz and 6.3\,GHz, respectively) and the rest, SNRs and background radio sources. %

\item The diffuse RC emission has two different structural components,  an extended structure with a spiral pattern covering parts of the optical arms and almost filling the inter-arm regions, and  constituting $\simeq 20\%$ of the total RC, and a diffuse disk (+halo)  component of $\simeq 40\%$ contribution. Both these components are dominated by the nonthermal synchrotron emission, particularly at the lower frequency of 1.5\,GHz.

\item Excluding the diffuse components, we find a linear correlation between non-radio SFR tracers and the observed RC emission, based on which radio SFR calibration relations are provided.

\item Both the mean nonthermal synchrotron spectral index obtained between 1.5 and 6.3\,GHz ($\alpha_{\rm n}\simeq -0.7$, $I_{\rm nt}\propto \nu^{\alpha_{\rm n}}$) and the mean equipartition magnetic field strength (B~$\simeq7\,\mu$G) agree with those obtained by \cite{Tabatabaei_2_07} and \cite{Tabatabaei_08} based on previous VLA observations. The VLA observations uncover variations in $\alpha_{\rm n}$ and B on spatial scales about 10 times smaller than observed before. We find that $\alpha_{\rm n}$ becomes flatter with  $\Sigma_{\rm SFR}$, indicating the escape and re-acceleration of CRes. The magnetic field strength also increases with   $\Sigma_{\rm SFR}$, following a bimodal relation, i.e., changing the slope above a SFR threshold.

\item Comparing the observed RC with the {molecular gas emission, we find a tighter correlation in star-forming regions than in other regions. No significant correlation is found between the molecular gas and thermal RC in the more quiescent and lower-density regions of the ISM. The nonthermal emission is, however, still correlated with the molecular gas in these regions}. The thermal RC--CO correlation in star-forming regions is in favor of a linear {Kennicutt-Schmidt (KS) star-formation law.} This is in contrast with previous studies in M\,33 where diffuse emission was included.

\end{itemize}

Similar studies in large samples of galaxies will shed light on the interplay between massive star formation and the ISM and its role in the evolution of SFR. The upcoming sensitive surveys with the Square Kilometer Array (SKA) and the ngVLA will enable mapping the RC emission from diffuse ISM not only in nearby galaxies but also at higher redshifts.

\section*{Acknowledgements}

J.H.K. acknowledges financial support from the State Research Agency
(AEI-MCINN) of the Spanish Ministry of Science and Innovation under the
grant "The structure and evolution of galaxies and their central
regions" with reference PID2019-105602GB-I00/10.13039/501100011033, from
the ACIISI, Consejer\'{i}a de Econom\'{i}a, Conocimiento y Empleo del
Gobierno de Canarias and the European Regional Development Fund (ERDF)
under grant with reference PROID2021010044, and from IAC project
P/300724, financed by the Ministry of Science and Innovation, through
the State Budget and by the Canary Islands Department of Economy,
Knowledge and Employment, through the Regional Budget of the Autonomous
Community. This research is partly based on observations
made with the WHT operated on the island of La Palma by the
Isaac Newton Group of Telescopes, in the Spanish Observatorio
del Roque de Los Muchachos of the Instituto de Astrof\'{i}sica de
Canarias.

\subsection*{Data Availability}
{The data underlying this article are partly available in its supplementary material. The VLA raw data are available on VLA archive under the project ID, 11B-145. The reduced/calibrated VLA data will be made available on CDS and/or NASA/IPAC Extragalactic Database. The combined VLA and Effelsberg maps will be shared on reasonable request to the corresponding author.}



\bibliographystyle{mnras}
\bibliography{s} 

\bsp	
\label{lastpage}
\end{document}